\documentclass{aa}  

\usepackage{graphicx}
\usepackage{placeins}
\usepackage{txfonts}
\usepackage{hyperref}
\hypersetup{
    colorlinks=true,
    linkcolor=blue,
    filecolor=magenta,      
    urlcolor=cyan,
    citecolor=blue
    }
\usepackage[dvipsnames]{xcolor}

\begin{document}

   \title{Cosmological constraints from the \textit{Planck} cluster catalogue with new multi-wavelength mass calibration from \textit{Chandra} and CFHT}

   \author{G. Aymerich\inst{1,2} \fnmsep\thanks{\email{gaspard.aymerich@universite-paris-saclay.fr}} \and
        M. Douspis\inst{1} \and
        G. W. Pratt\inst{2} \and
        L. Salvati\inst{1} \and
        E. Soubrié\inst{1} \and
        F. Andrade-Santos\inst{3,4} \and \\
        W. R. Forman\inst{3} \and 
        C. Jones\inst{3} \and 
        N. Aghanim\inst{1} \and
        R. Kraft\inst{3} \and
        R. J. van Weeren\inst{5} 
          }

   \institute{
         Université Paris-Saclay, CNRS, Institut d'Astrophysique Spatiale, 91405, Orsay, France
\and
         Université Paris-Saclay, Université Paris Cité, CEA, CNRS, AIM, 91191, Gif-sur-Yvette, France   
\and
         Center for Astrophysics $\vert$ Harvard \& Smithsonian, Cambridge, MA 02138, USA
\and
          Department of Liberal Arts and Sciences, Berklee College of Music, 7 Haviland Street, Boston, MA 02215, USA
\and
          Leiden Observatory, Leiden University, PO Box 9513, 2300 RA Leiden, the Netherlands
             }

   \date{Received XXX; accepted YYY}

  \abstract{We provide a new scaling relation between $Y_{\text{SZ}}$, the integrated Sunyaev-Zeldovich signal and $M_{500}^{Y_{\text{X}}}$, the cluster mass derived from X-ray observations, using a sample of clusters from the \textit{Planck} Early Sunyaev-Zeldovich (ESZ) catalogue observed in X-rays by \textit{Chandra}, and compare it to the results of the \textit{Planck} collaboration 
  obtained from \textit{XMM-Newton} observations of a subsample of the ESZ.
  We calibrated a mass bias on a subset of the \textit{Planck} cosmological cluster sample using published weak-lensing data from the Canadian Cluster Cosmology Project (CCCP) and Multi Epoch Nearby Cluster Survey (MENeaCS), 
  for the new scaling relation as well as that from the \textit{Planck} collaboration. We propose a novel method to account for selection effects and find a mass bias of $(1-b)=0.89\pm0.04$ for the \textit{Chandra}-calibrated scaling relation, and $(1-b)=0.76\pm0.04$ for the \textit{XMM-Newton}-calibrated scaling relation.
  We applied the scaling relations we derived to the full \textit{Planck} cosmological cluster sample and obtain new constraints on the cosmological parameters. 
  We find identical constraints regardless of the X-ray sample used,
  with $\sigma_8 = 0.77\pm0.02$, $\Omega_m = 0.31\pm0.02$, and $S_8 \equiv \sigma_8 \sqrt{\Omega_m / 0.3}=0.78\pm0.02$. We also provide constraints with a redshift evolution of the scaling relation fitted from the data 
  instead of fixing it to the self-similar value. We find a redshift evolution significantly deviating from the self-similar value, leading to a higher value of $S_8=0.81\pm0.02$. We compare our results to those 
  from various cosmological probes, and find that our $S_8$ constraints are competitive with the tightest constraints from the literature. When assuming a self-similar redshift evolution, our constraints 
  are in agreement with most late-time probes and in tension with constraints from the cosmic microwave background (CMB) primary anisotropies. When relaxing the assumption of redshift evolution and fitting it to the data, we find no significant tension with results 
  from either late-time probes or the CMB.}


    \titlerunning{New cosmological constraints from \textit{Planck}, \textit{Chandra} and the CFHT}
   \maketitle
%

\section{Introduction}
\label{intro}
In the hierarchical structure formation scenario, galaxy clusters emerge from peaks in the initial density field
and grow through accretion and mergers, due to the depth of their gravitational potential wells \citep{kravtsov_formation_2012}. Galaxy clusters are therefore tracers of the formation of cosmic structures throughout the evolution of the Universe 
and can be used to constrain the parameters of the cosmological model \citep{white_baryon_1993, henry_measurement_1997, allen_cosmological_2011}. 
In particular, the abundance of clusters as a function of redshift and mass is closely related to the matter density $\Omega_{\text{m}}$ and the amplitude of density fluctuations $\sigma_8$, and to the dark energy 
equation of state when considering extensions to the standard $\Lambda$-Cold Dark Matter ($\Lambda$CDM) cosmological model. Galaxy clusters are thus a powerful cosmological probe, 
and their abundance has been used to constrain cosmological parameters in numerous studies \citep[e.g.][]{vikhlinin_chandra_2009-1, rozo_cosmological_2010, planck_collaboration_planck_2014, bocquet_cluster_2019, costanzi_cosmological_2021, garrel_xxl_2022, lesci_amico_2022, bocquet_spt_2024, ghirardini_srgerosita_2024}.\\
A key feature of galaxy clusters is their multi-component nature: they are composed of a dark matter halo, which is the main component of their mass, 
and of baryonic matter in the form of both hot gas and galaxies. They can therefore be observed through multiple physical processes depending on the wavelength, revealing different properties of the cluster. 
In optical wavelengths, the member galaxies can be directly observed, and the cluster can also be detected through the gravitational lensing effect on background galaxies. In X-rays, the hot gas emits 
thermal bremsstrahlung radiation, and in the microwave wavelengths, the hot gas interacts with the cosmic microwave background (CMB) photons through the thermal Sunyaev-Zel'dovich (SZ) effect \citep{sunyaev_observations_1972}.\\
SZ surveys are a choice candidate for cluster cosmology since CMB experiments provide wide-field cluster surveys at no additional observational cost. 
Additionally, the SZ signal is independent of the cluster redshift, allowing for detections up to $z$$\sim$$2$ with the most recent surveys. Recently, SZ cluster catalogues have been obtained from the \textit{Planck} satellite 
\citep{planck_collaboration_planck_2014-1, planck_collaboration_planck_2016-1}, the South Pole Telescope \citep{bleem_galaxy_2015, bleem_sptpol_2020, klein_spt-sz_2024, bleem_galaxy_2024}, and the Atacama Cosmology Telescope \citep{hilton_atacama_2021}.\\
In order to compare cluster abundance observations with theoretical predictions, it is necessary to know the cluster mass, which is the fundamental quantity in the theory of structure formation \citep{pratt_galaxy_2019}.
While the SZ effect allows for the construction of large cluster catalogues, it does not provide a direct measurement of the cluster mass, as the observed quantity is the integrated SZ signal, 
noted $Y_{\text{SZ}}$ in this work, which is proportional to the gas pressure integrated along the line of sight. To use SZ surveys as a cosmological probe, it is thus necessary to relate the 
SZ signal to the cluster mass, usually via a scaling relation calibrated using a sub-sample of clusters from the SZ catalogue for which other observations allow for mass estimations. This mass calibration step is 
critical, as it is the main source of uncertainty and systematic error in the cosmological constraints obtained from SZ surveys. An example of the importance of mass calibration is found in 
\cite{planck_collaboration_planck_2016}, where constraints were obtained from the same SZ catalogue using three different mass calibrations varying by up to $\sim$$30 \%$, 
from Weighing the Giants \citep{von_der_linden_robust_2014}, the Canadian Cluster Comparison Project (CCCP) \citep{hoekstra_canadian_2015}, and CMB lensing \citep[calibration performed by the \textit{Planck} 
collaboration, using the method presented in][]{melin_measuring_2015}, yielding 
$\sigma_8 \left(\Omega_m / 0.31\right)^{0.3}$ values differing by more than 2$\sigma$, and tensions with the value obtained from the analysis of the primary anisotropies of the CMB ranging from nearly full agreement to 
$\sim$$2\sigma$.\\
Mass estimates can be obtained from X-ray observations that provide precise gas density and temperature profiles, from which masses can be calculated under the hydrostatic equilibrium hypothesis 
, that is, assuming that the gas contained in the cluster is at equilibrium, with the thermal pressure balancing the gravitational potential. Mass estimates from X-ray observations have low statistical uncertainties 
but are subject to systematic errors. These errors can be due to the presence of non-thermal pressure, coming from magnetic fields \citep[e.g.][]{dolag_effect_2000}, 
cosmic rays \citep[e.g.][]{boss_span_2022}, or turbulence \citep[e.g.][]{rasia_dynamical_2004, pearce_hydrostatic_2020, gianfagna_exploring_2021}. Another source of error in mass determination is the dynamical 
state and local environment of the cluster \citep{gouin_shape_2021,gouin_gas_2022}. Complex temperature structures in the intra-cluster medium can also affect the mass calculated from X-ray observations \citep{rasia_systematics_2006}.\\ 
The mass of a cluster can also be estimated through the gravitational lensing effect, which does not require assumptions on the dynamical state of the cluster, but usually suffers from 
large statistical uncertainties due to the long exposure times required to get a sufficient number of background galaxies.\\
In the \textit{Planck} collaboration 2013 and 2015 papers \citep{planck_collaboration_planck_2014,planck_collaboration_planck_2016}, cosmological parameters were constrained using the number counts of SZ-detected clusters. 
In this work, our aim is to improve on the \textit{Planck} collaboration approach by focusing on the mass calibration step. We leverage a larger, SZ-selected X-ray calibration sample from the \textit{Chandra-Planck} Legacy Program 
as well as a larger weak-lensing sample from \cite{herbonnet_cccp_2020} using data from the Canadian Cluster Comparison Project (CCCP) and the Multi Epoch Nearby Cluster Survey (MENeaCS) to provide a new mass calibration and cosmological constraints.\\
In addition to the improved statistical power, our results provide a robustness check regarding potential instrumental systematics to the constraints formerly obtained, 
as the \textit{Planck} X-ray calibration sample was obtained from \textit{XMM-Newton} data, whereas the new calibration sample was obtained from \textit{Chandra} observations of \textit{Planck}-detected clusters. Given that the two 
instruments are known to yield different results \citep[see][]{schellenberger_xmm-newton_2015, potter_hydrostatic_2023}, evaluating the effect of instrumental differences on the final constraints is crucial to understanding the reliability of our results.\\
The paper is organised as follows: in Sect. \ref{data}, we present the data sets used in this work, and in Sect. \ref{scaling_relation}, we calibrate the $Y_{\text{SZ}}-M_{500}$ scaling relation using the \textit{Chandra-Planck} sample.
In Sect. \ref{cosmo}, we constrain the cosmological parameters using the \textit{Planck} cosmological cluster sample and the scaling relation calibrated in Sect. \ref{scaling_relation}, and compare our results with the ones obtained by the \textit{Planck} collaboration.
In Sect. \ref{discussion}, we discuss the possible systematics impacting our cosmological constraints, explore the impact of fitting the redshift dependence of the scaling relation instead of fixing it to the self-similar value, and compare our results to the ones from recent analyses based on various cosmological probes.
Finally, in Sect. \ref{conclusion}, we summarise our results and conclude.\\
For this work, we use the following reference cosmology: $\Omega_m=0.3$, $\Omega_\Lambda=0.7$, $H_0=70$ km s$^{-1}$ Mpc$^{-1}$. Unless explicitly stated, all quantities are given within $R_{500}$, 
the radius within which the mean density is 500 times the critical density of the Universe at the cluster redshift.

\section{Data and analysis}
\label{data}

\subsection{\textit{Chandra} observation of \textit{Planck} ESZ sample}
\label{Chandra_Planck}
\begin{figure}[]
  \centering
  \includegraphics[width=\columnwidth]{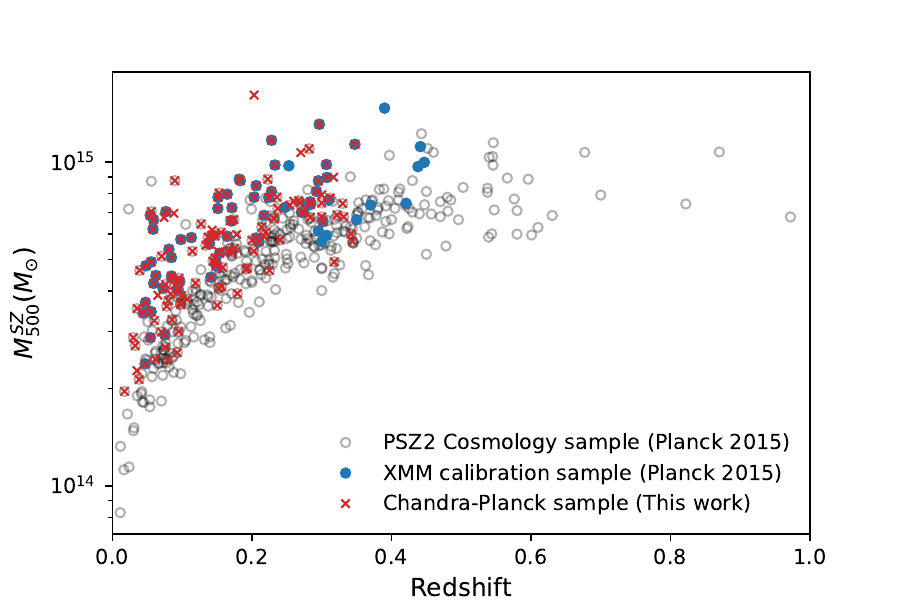}
  \caption{Mass-redshift distribution of the \textit{Chandra-Planck} sample, the \textit{Planck} MMF3 cosmological sample, and the \textit{XMM-Newton} calibration sample used by the \textit{Planck} collaboration.}
  \label{fig:M_z_cosmo_XMM_Chandra}
  \end{figure}

For our investigation, we use the data set from the \textit{Chandra-Planck} Legacy Program for Massive Clusters of Galaxies\footnote{\url{https://hea-www.cfa.harvard.edu/CHANDRA_PLANCK_CLUSTERS/}}, 
which is a full follow-up observation with the \textit{Chandra} X-ray observatory of the 163 clusters with redshift $z<$0.35 from the \textit{Planck} ESZ catalogue, which is constructed from 
clusters detected with a signal-to-noise ratio above 6 in the \textit{Planck} early results maps \citep{planck_collaboration_planck_2011}. All 163 clusters were observed with sufficient exposures to obtain a 
minimum of 10000 source counts. Six clusters that fall outside the more restrictive mask used to obtain the \textit{Planck} PSZ2 cosmological catalogue \citep{planck_collaboration_planck_2016-1} were discarded from the analysis: 
PLCKESZ G115.71+17.52, G228.49+53.12, G269.51+26.42, G275.21+43.92, G282.49+65.17, G315.70-18.04.
Nine clusters were identified as multiple systems in the X-ray data, and were also discarded as the size of the \textit{Planck} beam did not allow for proper disentanglement of the two components in the SZ data: 
PLCKESZ G039.85-39.98, G062.42-46.41, G107.11+65.31, G124.21-36.48, G195.62+44.05, G241.85+51.53, G263.20-25.21, G337.09-25.97, G345.40-39.34. 
Finally, we removed PLCKESZ G234.59+73.01 (A1367/Leo cluster at z = 0.02) and PLCKESZ G057.33+88.01 
(A1656/Coma cluster at z = 0.02), two nearby clusters whose size did not allow for a proper data reduction. The final sample thus contains 146 clusters with high-quality data in both X-ray and SZ, 
allowing for a multi-wavelength analysis and robust calibration of the $Y_{\text{SZ}}-M_{500}$ scaling relation.\\
The sample and X-ray data reduction process is fully described in \cite{andrade-santos_chandra_2021} \citep[see also][]{vikhlinin_chandra_2005}, and we refer the reader to these papers for more details. 
A summary of the data reduction process is given in Sect. \ref{processing}.\\
Fig. \ref{fig:M_z_cosmo_XMM_Chandra} shows the redshift and mass distribution of the sample used in this work, and compares it to the \textit{Planck} PSZ2 cosmological sample and 
the X-ray calibration sample used in \cite{planck_collaboration_planck_2014} and \cite{planck_collaboration_planck_2016-1}.\\

\subsection{Data sets of the \textit{Planck} collaboration}
The \textit{Planck} PSZ2 cosmological sample, presented in \cite{planck_collaboration_planck_2016-1}, is composed of all the clusters detected in 
\textit{Planck} HFI maps by the matched multi-frequency filter algorithm MMF3 \citep{melin_catalog_2006} with a signal-to-noise ratio above 6. 
The detection procedure uses a more restrictive mask of contaminated regions than the full PSZ2 catalogue, leaving only 65\% of the sky unmasked, 
to ensure a very high purity of the sample.
It contains 439 clusters and is the sample used to constrain cosmological parameters via MCMC fitting 
of the number counts in \cite{planck_collaboration_planck_2016}. To estimate the masses of the clusters, the \textit{Planck} collaboration used a scaling relation between the SZ signal and the cluster mass, 
calibrated on a sub-sample of 71 clusters with X-ray observations from \textit{XMM-Newton} \cite{planck_collaboration_planck_2014}. 
In this work, we use a new calibration sample of 146 clusters from the \textit{Planck} ESZ catalogue with \textit{Chandra} X-ray observations.\\
Compared to the \textit{XMM-Newton} sample, the \textit{Chandra} calibration sample has more 
clusters and extends to lower masses and redshifts. Given the cut at redshift 0.35, the maximum redshift of the \textit{Chandra} sample is lower than the one of the \textit{XMM-Newton} sample, which has six clusters with $0.35<z<0.45$.\\
Beyond the mass and redshift ranges, the most important difference between the two calibration samples is the selection of the sample. As a full X-ray follow-up of the \textit{Planck} ESZ sample, the \textit{Chandra} sample 
is SZ-selected (with the additional removal of the nine multiple systems and two nearby clusters) and therefore is subject to the same type of selection biases as the clusters in the 
\textit{Planck} catalogues. The \textit{XMM-Newton} sample, on the other hand, is a subset of 71 detections from the \textit{Planck} cosmological cluster sample, detected
at signal-to-noise ratio above 7, for which good quality \textit{XMM-Newton} observations were available \citep{planck_collaboration_planck_2014}. The \textit{XMM-Newton} sample is therefore not 
guaranteed to be representative of the \textit{Planck} cosmological sample and is likely to be biased towards X-ray bright clusters. Since the selection function of the X-ray sample is not known, 
it was not possible to correct for this bias in \cite{planck_collaboration_planck_2014}. This issue of the representativity of the calibration sample is investigated in depth in Appendix \ref{representativity}.\\

\subsection{Weak lensing data: CCCP and MENeaCS samples}
\label{weak_lensing}
In this work, we use the results of \cite{herbonnet_cccp_2020} who derived masses for 100 clusters from the Canadian Cluster Comparison Project (CCCP) and the Multi Epoch Nearby Cluster Survey (MENeaCS). 
It currently is one of the largest coherent weak-lensing cluster samples, combining two data sets coming from observations with the Canada-France-Hawaii Telescope (CFHT) with similar depth, and analysing them with the same pipeline. A detailed description of the CCCP sample is available in \cite{hoekstra_canadian_2012} and \cite{sand_intracluster_2011, sand_multi-epoch_2012} provide a full description of the MENeaCS sample. Both data sets are X-ray-selected cluster samples, with MENeaCS containing 48 low redshift clusters with $0.05<z<0.15$ and CCCP containing 52 higher redshift clusters with $0.15<z<0.55$. The analysis procedure used to derive the masses from the optical imaging surveys is described in \cite{herbonnet_cccp_2020}.

\subsection{X-ray data reduction and mass estimates}
\label{processing}
The full data reduction process from raw event files to profiles is described in \cite{andrade-santos_chandra_2021}. In this section, we briefly summarise the main steps of the data reduction process.\\
Charge-transfer inefficiency, mirror contamination, CCD non-uniformity, and time dependence of gain are corrected for during data reduction. High background periods are removed, and blank sky background and readout artefacts are subtracted from the signal. 
Point sources and extended substructures are masked before analysing the cluster emission.\\
Surface brightness profiles are extracted in the 0.7-2 keV band, where the signal-to-noise ratio is maximal, in concentric annuli around the X-ray emission peak. Spectra are extracted in larger concentric annuli, 
and fitted with an absorbed single-temperature thermal model. The projected temperature 
profiles are then fitted with a 3D temperature model to obtain the deprojected temperature profiles. These are then used to compute the emission measure profiles 
from the surface brightness, which are fitted with the gas density model given by \cite{vikhlinin_chandra_2006}.\\
In \cite{andrade-santos_chandra_2021}, total cluster masses are computed from the gas mass and temperature, using the $Y_\text{X}-M_{500}$ scaling relation calibrated in \cite{vikhlinin_chandra_2009}:
\begin{equation}
  \label{eq:Y_X-M_500}
   \begin{array}{l}
       M_{500} = E(z)^{-2/5} A_{\text{YM}} \left(\frac{Y_\text{X}}{3 \, \cdot 10^{14} \, M_\odot\text{keV}}\right)^{B_{\text{YM}}},
   \end{array}
\end{equation}
where $A_{\text{YM}}=(5.77 \pm 0.20) \cdot 10^{14} \, h^{1/2} \, M_\odot$, $B_{\text{YM}}={0.57 \pm 0.03}$, and $E(z)=\sqrt{\Omega_m(1+z)^3+\Omega_\Lambda}$.\\
The quantity $Y_{\text{X}} = M_\text{gas} T_\text{X}^\text{exc}$ is defined in \cite{kravtsov_new_2006} as the product of $M_\text{gas}$, the gas mass inside $R_{500}$, and $T_\text{X}^\text{exc}$, the core-excised temperature obtained by fitting a single temperature MEKAL model to the spectral data in the $0.15 R_{500}$ to $R_{500}$ region. 
It is thus measured within $R_{500}$, and the mass calculations have to be done following an iterative process. 
A first mass estimate is computed from the $T_\text{X}-M_{500}$ scaling relation, giving a first $R_{500}$ value. The $Y_{\text{X}}$ value is then computed within this first $R_{500}$, which gives a new mass
estimate, leading to new $R_{500}$ and $Y_{\text{X}}$ values, and so on until convergence. At each step, to calculate $Y_{\text{X}}$, $T_\text{X}^{exc}$ is re-extracted in the new $0.15 R_{500}$ to $R_{500}$ region.\\
A possible caveat of the approach is that the \cite{vikhlinin_chandra_2009} scaling relation was derived for an X-ray-selected sample, whereas the sample discussed in this work is SZ-selected. However, there is currently no strong evidence that the $Y_\text{X}-M_{500}$ scaling relations recovered from X-ray and SZ-selected samples differ. Theoretically, the original motivation for $Y_\text{X}$ was that it is a quantity that is independent of selection and/or dynamical state \citep[as is detailed in][]{kravtsov_new_2006}. Observationally, Fig. 8 of \cite{andrade-santos_chandra_2021} compares the $Y_\text{SZ}-Y_\text{X}$ relations for an SZ- and an X-ray selected sample. There clearly is no evidence for any difference between the SZ and the X-ray-selected systems. We therefore assume that the scaling relation of Eq. \ref{eq:Y_X-M_500} is valid for our sample.

\subsection{SZ signal extraction from \textit{Planck} data}
\label{SZ_data}
As shown in \cite{planck_collaboration_planck_2011-2}, using an external prior on the size and position can lead to more robust SZ flux estimates, given the size of the \textit{Planck} beam (around 7 arcmin).
To calibrate the $Y_{\text{SZ}}-M_{500}$ scaling relation, we use the SZ signal extracted from the \textit{Planck} DR2 maps, by running a matched multi-frequency filter algorithm \citep{herranz_scaleadaptive_2002}, with the X-ray derived positions 
and sizes as prior information. 
We use the \textit{Planck} DR2 maps, even though newer maps are available as this work uses the PSZ2 cosmological sample obtained on DR2 maps. 
While deriving a new catalogue from the most recent \textit{Planck} data would be possible and possibly beneficial in terms of final cosmological constraints, 
it would introduce an unnecessary inconsistency in the comparison of mass calibration samples between this work and the original \textit{Planck} 2015 analysis. 
The MMF algorithm extracts the $Y_{\text{SZ}}$ signal from the six \textit{Planck} HFI maps (100, 143, 217, 353, 545, and 857 GHz) using the SZ frequency spectrum and the universal pressure profile from \cite{arnaud_universal_2010}, 
scaled to an aperture $\theta_{500}$ corresponding to $R_{500}^{\text{X-ray}}$, convolved with the instrumental beam as a template. The extraction is performed on a 10°$\times$10° patch centred on the X-ray cluster position, 
with a pixel size of 1.72$\times$1.72 arcmin$^2$. The signal is integrated up to $5R_{500}$, then scaled back to $R_{500}$, using the assumed pressure profile. The noise auto- and cross-spectra are directly calculated from the data. 
This python MMF implementation uses the same approach and steps as the IDL MMF3 implementation used in \cite{planck_collaboration_planck_2016-1}. We refer the reader to that paper, and references therein, for more details.\\
This extraction step was done both in \cite{andrade-santos_chandra_2021} and in this work, using the same X-ray input positions and aperture, but fully independently developed MMF algorithms. Appendix \ref{MMF_comparison} 
shows the excellent agreement between the results.

\section{$Y_{\text{SZ}}-M_{500}$ scaling relation}
\label{scaling_relation}
Relating the SZ signal to the cluster mass is required to compare the cluster number counts 
with theoretical predictions and constrain the underlying cosmological parameters. In this section we calibrate a $Y_{\text{SZ}}-M_{500}$ scaling relation, using the \textit{Chandra} calibration sample described 
in Sect. \ref{data}, for which we have both SZ measurements and mass estimates from the X-ray observations. We also use the weak-lensing masses derived in \cite{herbonnet_cccp_2020} as a true mass anchor 
to calibrate the hydrostatic mass bias.\\
\subsection{Calibrating the scaling relation}
\label{calibration_scaling_relation}
\begin{figure}[h]
  \centering
  \includegraphics[width=\columnwidth]{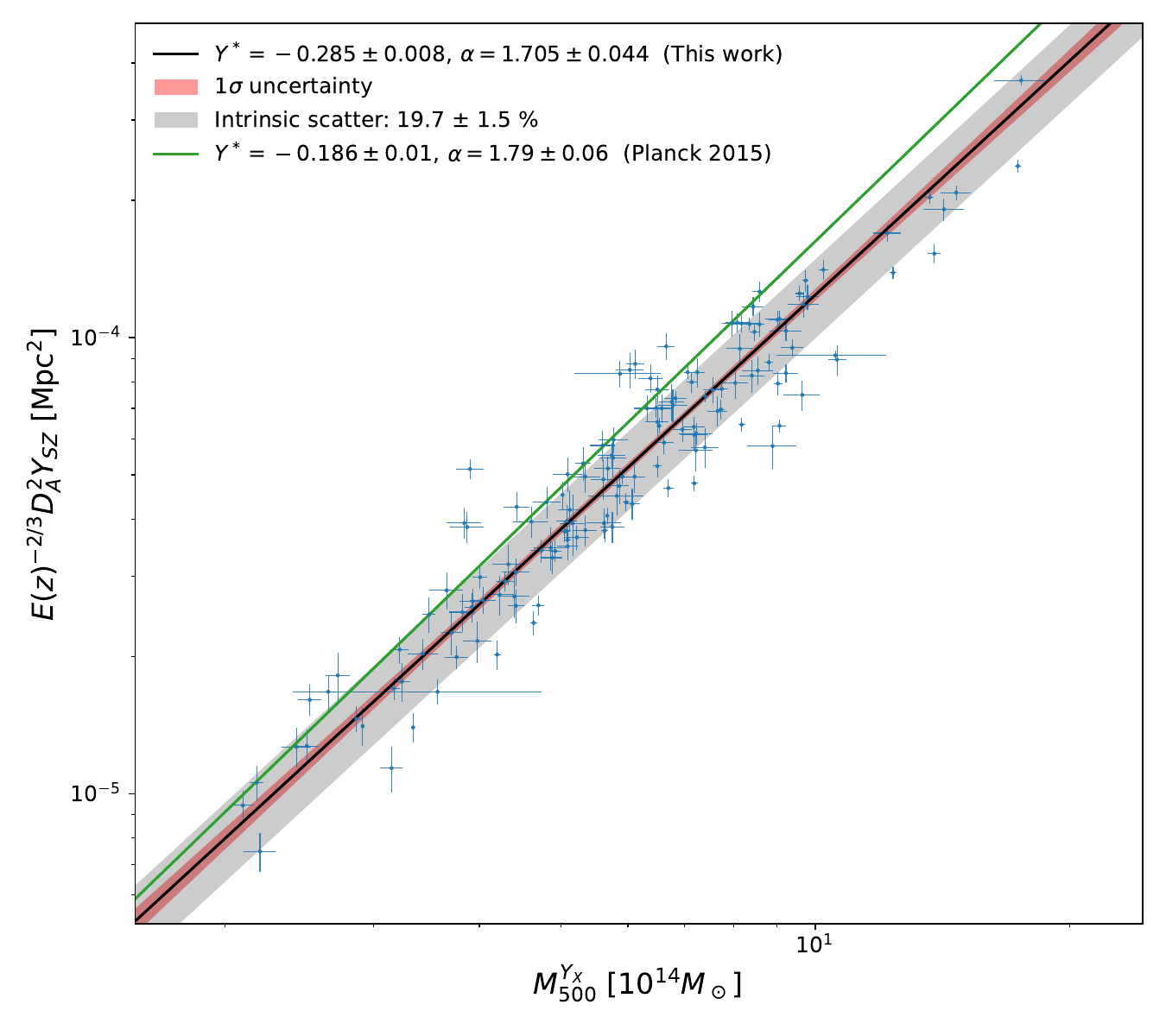}
  \caption{Calibration of the $Y_{\text{SZ}}-M_{500}^{Y_{\text{X}}}$ scaling relation. The blue dots correspond to the X-ray mass and SZ signal (with $1 \sigma$ uncertainties) for each cluster in the sample. The black line shows the best-fit relation obtained in this work, with the red regions corresponding to $1\sigma$ uncertainties,
  the grey regions to the intrinsic scatter, and the green line to the best-fit relation obtained in \cite{planck_collaboration_planck_2014}.}
    \label{fig:M_Y_sz_scaling_Santos}  
  \end{figure}

To study the effect of changing the calibration sample, we calibrate the $Y_{\text{SZ}}-M_{500}$ scaling relation following the \textit{Planck} collaboration approach described in appendix A 
of \cite{planck_collaboration_planck_2014}. Since the ESZ sample is signal-to-noise limited, the detected objects are biased high with respect to the mean close to the signal-to-noise threshold, due to the intrinsic scatter of the relation. 
This effect, commonly known as Malmquist bias, needs to be corrected to retrieve the underlying scaling relation in an unbiased way. We follow the same approach as in \cite{planck_collaboration_planck_2014}, which is based on the method presented in \cite{vikhlinin_chandra_2009}, to account for Malmquist bias. The detailed justification of the method can be found in Appendix A.2 of \cite{vikhlinin_chandra_2009}, simply replacing luminosity with $Y_{\text{SZ}}$ and the flux cut with a signal-to-noise cut. Each individual $Y_{\text{SZ}}$ value is corrected by
dividing it by the mean bias $m$ at the corresponding signal-to-noise ratio, assuming a log-normal intrinsic scatter $\sigma_\text{int}$:
\begin{equation}
  \begin{array}{l}
    Y_{\text{SZ}}^{\text{corrected}}=Y_{\text{SZ}}/m \: \text{ with } \: \text{ln} \, m=\frac{\text{exp}\left(-x^2/2\sigma^2\right)}{\sqrt{\pi/2} \, \text{erfc}\left(-x/\sqrt{2}\sigma\right)}\sigma,
  \end{array}
\end{equation}
where $(S/N)_\text{cut}$ is the signal-to-noise threshold of the sample, $x=-\text{log}\left(\frac{(S/N)}{(S/N)_\text{cut}}\right)$, and $\sigma=\sqrt{\text{ln}[((S/N)+1)/(S/N)]^2+(\text{ln}\,10 \, \sigma_\text{int})^2}$.\\
The corrected $Y_{\text{SZ}}$ values are then used to calibrate the $Y_{\text{SZ}}-M_{500}^{Y_{\text{X}}}$ scaling relation, assuming the following form for the relation:\\
\begin{equation}
  \label{eq:Y_SZ-M_500_form}
  E^{-2/3}(z)\frac{D^2_A\,Y_{\text{SZ}}}{Y_{\text{piv}}}=10^{Y^*}\left(\frac{M_{500}^{Y_{\text{X}}}}{M_{\text{piv}}}\right)^{\alpha}.
\end{equation}
To allow for easy comparison, the pivot points were chosen to be the same in this work as in \cite{planck_collaboration_planck_2014}, that is, $M_{\text{piv}}=6\cdot10^{14}M_\odot$ and $Y_{\text{piv}}=10^{-4}\text{Mpc}^2$.
We fit the data via a Markov Chain Monte Carlo (MCMC) approach with the \texttt{emcee} sampler, using a standard Gaussian log-likelihood with intrinsic scatter:
\begin{equation}
  \label{eq:likelihood}
  \begin{array}{l}
  \text{ln}\mathcal{L}=-0.5\sum_{i=1}^{N}\left[\frac{(y_i-(\alpha x_i+10^{Y^*}))^2}{2\sigma_i^2}-\text{ln}\left(\frac{1+10^{2 Y^*}}{2\pi\sigma_i^2}\right)\right],
  \end{array}
\end{equation}
where $x_i$ and $y_i$ are the logarithm of the mass and $Y_{\text{SZ}}$ values of the $i^{th}$ cluster, $\sigma_{x_{i}}$ and $\sigma_{y_{i}}$ are the uncertainties on the mass and $Y_{\text{SZ}}$ values, 
$\sigma_\text{int}$ is the intrinsic scatter of the relation, and $\sigma_i^2=\sigma_\text{int}^2+10^{2 Y^*} \sigma_{x_{i}}^2+\sigma_{y_{i}}^2$.\\
The varying parameters are the normalisation $Y^*$, the slope $\alpha$, and the intrinsic scatter $\sigma_\text{int}$. The only prior is requiring the 
intrinsic scatter to be positive. Since the intrinsic scatter is not known a priori and is required to calculate the Malmquist bias correction, this fitting procedure is an iterative process, where the intrinsic scatter used to compute the mean bias is updated at each step to the value obtained once the relation is fitted. Since the Malmquist bias correction is negligible for most clusters, the convergence is reached after only 4 to 5 steps.\\
Two other linear regression methods were tried: LinMix \citep{kelly_aspects_2007} and BCES \citep{akritas_linear_1996}, with no significant difference in the results. 
We thus only report the results obtained with the MCMC method using the \texttt{emcee} sampler in the rest of this work.
Fig. \ref{fig:M_Y_sz_scaling_Santos} shows the scaling relation obtained with the \texttt{emcee} sampler, and compares it with the one obtained by the \textit{Planck} collaboration
with the \textit{XMM-Newton} calibration sample. 
\\
This scaling relation is not final: the $Y_{\text{SZ}}-M_{500}^{Y_{\text{X}}}$ relation needs to be combined with the $M_{500}^{Y_{\text{X}}}-M_{500}$ relation to obtain the $Y_{\text{SZ}}-M_{500}$ relation.
The $M_{500}^{Y_{\text{X}}}-M_{500}$ relation has the following form, with $\sigma_A$ and $\sigma_{\alpha}$ the uncertainties in the normalisation and slope of the $Y_{\text{X}}-M_{500}$ relation taken from \cite{vikhlinin_chandra_2009}:
\begin{equation}
  \label{eq:M-M_Y_X}
  M_{500}^{Y_{\text{X}}} = 10^{\pm \sigma_A / \alpha} [(1-b)M_{500}]^{1 \pm \sigma_{\alpha} / \alpha}.
\end{equation}
Combining the $Y_{\text{SZ}}-M_{500}^{Y_{\text{X}}}$ and $M_{500}^{Y_{\text{X}}}-M_{500}$ relations, we obtain the following $Y_{\text{SZ}}-M_{500}$ relation:
\begin{equation}
  \label{eq:Y_SZ-M_500_Chandra}
  E^{-2/3}(z)\frac{D^2_A\,Y_{\text{SZ}}}{Y_{\text{piv}}}=10^{-0.29\pm0.01}\left(\frac{(1-b)M_{500}}{M_{\text{piv}}}\right)^{1.70\pm0.1}.
\end{equation}
Moving from $Y_{\text{SZ}}-M_{500}^{Y_{\text{X}}}$ to $Y_{\text{SZ}}-M_{500}$, the best-fit parameters do not change, but the uncertainties of the $M_{500}^{Y_{\text{X}}}-M_{500}$ relation are propagated, leading to increased uncertainties in the final $Y_{\text{SZ}}-M_{500}$ relation, 
by $\sim$$2$ for the slope and $\sim$$1.3$ for the normalisation. 
In addition, the scatter around the mean relation is also affected: assuming the scatter around both relations is independent, our final estimate of the scatter is $\sigma_{Y_{\text{SZ}}-M_{500}}=21\%$.

\subsection{Hydrostatic mass bias}
\label{bias}
\begin{figure}[]
    \centering
    \includegraphics[width=\columnwidth]{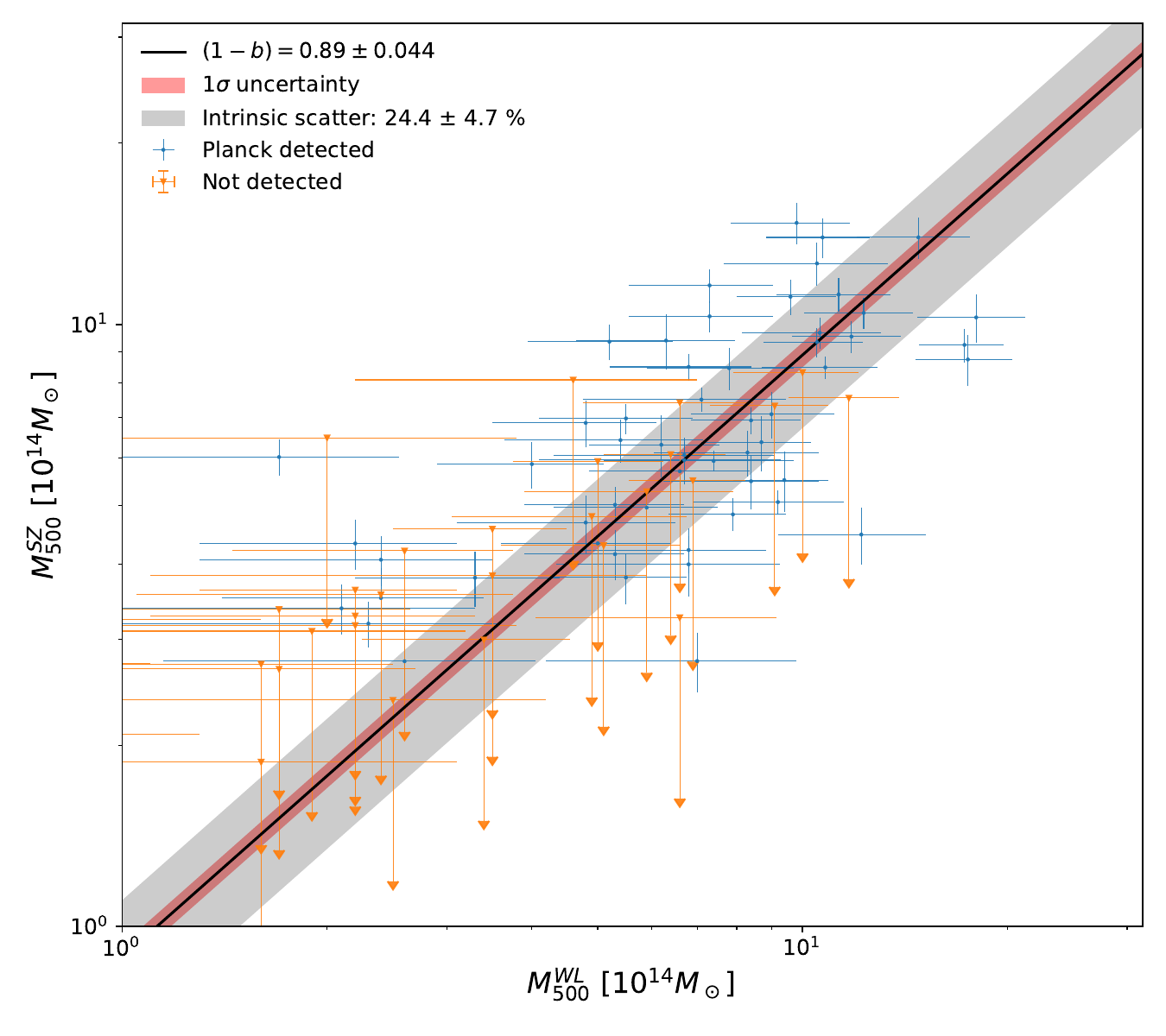}
    \caption{Hydrostatic mass bias determination for \textit{Chandra} scaling relation, with selection bias accounted for. The clusters detected by \textit{Planck} are plotted in blue, while the orange points are 
    the mass upper limits for the non-detected clusters. The black line shows the best-fit relation, with red contours corresponding to $1\sigma$ uncertainties, 
    and grey contours to the intrinsic scatter.}
    \label{fig:Mass_bias_Chandra_Malmquist}
\end{figure}

In the previous section, we calibrated the scaling relation between the SZ signal and the cluster mass. However, the masses used for calibration were obtained from X-ray observations and are 
not the true masses of the clusters, 
but biased estimates, since their calculation indirectly relies on the hypothesis of hydrostatic equilibrium, through the $Y_{\text{X}}-M_{500}$ relation. 
This is usually referred to as the hydrostatic mass bias and is accounted for by a multiplicative factor $(1-b)$ 
in the scaling relation \ref{eq:Y_SZ-M_500_Chandra}. This free parameter needs to be constrained using an independent, ideally bias-free, mass estimate of the clusters, before using the scaling relation 
to constrain cosmological parameters. In this work, we follow the approach of 
\cite{planck_collaboration_planck_2016} to use an external weak-lensing data set to constrain the bias. Similarly to \cite{planck_collaboration_planck_2016}, we consider a constant bias, 
but some studies have investigated a potential dependence of the bias on the cluster mass and redshift \citep[e.g.][]{salvati_mass_2019, wicker_constraining_2023}.
In this work, we use weak-lensing masses obtained in \cite{herbonnet_cccp_2020} to constrain the value of $(1-b)$.
This data set was already used to provide tight constraints on the hydrostatic mass bias in \cite{herbonnet_cccp_2020}. They compared 
the weak-lensing mass estimates with SZ masses derived from \textit{Planck} data for the 61 clusters in their data set that were detected by \textit{Planck}, and found a bias of $(1-b)=0.84\pm0.04$.\\
However, this bias value cannot be directly used in this work, since it was obtained by comparing the weak-lensing masses with the \textit{Planck} SZ-derived masses calculated using the scaling relation calibrated in 
\cite{planck_collaboration_planck_2014} using \textit{XMM-Newton} X-ray data. The value of the bias thus needs to be re-estimated using the new scaling relation calibrated in this work.
\\
\cite{battaglia_weak-lensing_2016} showed that selection effects can lead to overestimating the mass bias, with an expected effect between 3-15\%. Here, we propose a method to account for this effect, 
based on the idea proposed in \cite{battaglia_weak-lensing_2016} to treat the \textit{Planck} measurements as follow-up observations of the weak-lensing sample. When approaching the problem under this angle, 
a cluster from the weak-lensing sample not detected by \textit{Planck} is a source of information, since the \textit{Planck} catalogue is signal-to-noise-limited.\\
The first step is to remove the clusters of the weak-lensing sample that lie in the masked region of the \textit{Planck} data, since they will necessarily not be detected by \textit{Planck}, and will not provide any information. 
We find 8 clusters (A780, A1285, A2050, ZWCL1023, ZWCL1215, A2104, A2163, MS0440) in the mask used for the extraction of the PSZ2 cosmological catalogue, and remove them from the sample, after manually verifying their absence from the cosmological sample.
This leaves 92 clusters in the weak-lensing sample that can potentially have a match in the \textit{Planck} catalogue. We also remove 5 clusters that are part of multiple systems detected as a single source by \textit{Planck}: 
A115N, A115S, A223N, A223S, and A119 \citep[not removed in][]{herbonnet_cccp_2020}. 
We then match the remaining clusters with the full \textit{Planck} PSZ2 cosmological catalogue and 
find 56 direct matches, as opposed to 61 in \cite{herbonnet_cccp_2020} (60 matches when excluding A119). Since the list of matches is not available in \cite{herbonnet_cccp_2020}, our best guess is that the difference can be explained by 
the mask used, since they only remove 1 out of 100 clusters before matching due to lying in the \textit{Planck} mask, while we find 8 clusters in the mask.\\
We then calculate the \textit{Planck} mass of the matched clusters with the $Y_{\text{SZ}}-M_{500}$ scaling relation calibrated in this work:
\begin{equation}
  \label{eq:M_500_SZ}
  M_{500}^{\text{SZ}}=M_{\text{piv}} \left(10^{-Y^*} E^{-2/3}(z) \frac{D^2_A\,Y_{\text{SZ}}}{Y_{\text{piv}}}\right)^{\frac{1}{\alpha}}.
  \end{equation}
To break the size-flux degeneracy, we need to combine it with the $\theta_{500}-M_{500}$ relation \citep[see][]{planck_collaboration_planck_2016-1}. We use an MCMC approach to compute the masses, 
allowing for proper marginalisation over both the posterior probability distribution in the $Y_{\text{SZ}}-\theta_{SZ}$ plane given in the \textit{Planck} catalogue and the uncertainties of the scaling relation parameters 
(see Appendix \ref{sz_mcmc}).\\
If no match is found, we compute the upper limit of the \textit{Planck} mass given the $S/N=6$ detection threshold of the \textit{Planck} cosmological catalogue, 
using the noise level $N$ of the \textit{Planck} map at the cluster position, for the aperture $\theta_{500}$ corresponding to $R_{500}^{\text{WL}}$:
\begin{equation}
  \label{eq:M_500_SZlim}
  M_{500}^{SZ,\text{lim}}=M_{\text{piv}} \left(10^{-Y^*} E^{-2/3}(z) \frac{D^2_A\, 6 N}{Y_{\text{piv}}}\right)^{\frac{1}{\alpha}}.
  \end{equation}
The choice of the WL aperture is an approximation, but it is justified by the shallow dependence of the noise level on the aperture and the fact that $\theta_{500}$ scales as $M_{500}^{1/3}$.\\
We then use a modified version of the MCMC power-law regression algorithm detailed in Sect. \ref{scaling_relation} on the $M_{500}^{\text{SZ}} - M_{500}^{\text{WL}}$ data, 
with the slope set to unity, to find the best-fit normalisation value, which corresponds to the best fit hydrostatic mas bias value $(1-b)$. The likelihood is the same as in Eq.\ref{eq:likelihood} with $\alpha$ set to 1, 
and the addition of the method detailed below to deal with non-detections. For non-detected clusters, at each step of the MCMC algorithm, we draw a mass from a log-normal distribution, with the current model value at the WL mass as mean 
and standard deviation $\sigma^2=\sigma_{int}^2+\beta^2 \sigma_{x_{i}}^2+\sigma_{y_{i}}^2$. 
If the drawn mass is below the detection threshold, we keep the mass and set the uncertainty to the median relative uncertainty on the $M_{500}^{\text{SZ}}$ measurement of the detected clusters ($\sim$8\%). 
While this is a simplified assumption, it is justified by the fact that the scaling relation uncertainties dominate the total error budget, leading to a low standard deviation of the relative uncertainty on 
the final mass ($\sim$$2\%$). If the mass is above the threshold, we reject it and draw a new one.\\
We calculate the bias for the scaling relations calibrated with both \textit{Chandra} and \textit{XMM-Newton}, with and without including non-detected clusters (abbreviated as D+nD and D), to verify the robustness of our method. 
The fit for the \textit{Chandra} scaling relation, including non-detections, is shown in Fig. \ref{fig:Mass_bias_Chandra_Malmquist}, 
and Table \ref{table:bias} shows the results for all four configurations.
\begin{table}[!ht]
  \caption{Hydrostatic mass biases for both scaling relations.}             
  \label{table:bias}      
  \centering                          
  \begin{tabular}{c c c}        
  \hline\hline                 
  Calibration sample & D+nD & D \\    
  \hline                        
    \textit{Chandra} & $0.89 \pm 0.04$ & $0.91 \pm 0.05$ \\      
    \textit{XMM-Newton} & $0.76 \pm 0.04$ & $0.78 \pm 0.04$ \\
     
  \hline                                   
  \end{tabular}
  \tablefoot{The table presents hydrostatic mass biases $(1-b)$ calculated for both \textit{Chandra} and \textit{XMM-Newton} calibrated scaling relations, with (D+nD) or without (D) accounting for non-detected clusters.}
\end{table}
Our results are not fully coherent with the value of $(1-b)=0.84\pm0.04$ obtained in \cite{herbonnet_cccp_2020} for the \textit{XMM-Newton} calibrated scaling relation without accounting for the non-detected clusters. 
The difference can be attributed to a slightly different sample (we cannot reproduce four of the cluster matches), a different procedure of SZ mass estimation, and a different fitting method.\\
When using the scaling relation calibrated with \textit{Chandra} data, we find a higher $(1-b)$ value, as expected given the fact that cluster masses computed from \textit{Chandra} data are systematically 
higher than the ones computed from \textit{XMM-Newton} data, as discussed in Sect. \ref{comparison_scaling}. With both scaling relations, we find a 2-3\% lower $(1-b)$ value when accounting for the non-detected clusters, 
which is in agreement with the lowest estimations from \cite{battaglia_weak-lensing_2016}. This is expected since the results from \cite{battaglia_weak-lensing_2016} were obtained by setting the mass 
of the non-detected clusters either to the detection threshold, leading to a 3\% lower bias, or to zero, leading to a 15\% lower bias. Our method of imposing an upper limit at the detection threshold is 
closer to the first scenario investigated in \cite{battaglia_weak-lensing_2016}, and obtaining similar results is thus expected.\\
Unless explicitly stated, we use the $(1-b)$ values obtained including non-detected clusters in the rest of this work.
\subsection{Comparison with \textit{Planck} 2015}
\label{comparison_scaling}
Table \ref{table:scaling_relation} compares the scaling relation obtained in this work (see Eq.\ref{eq:Y_SZ-M_500_Chandra}), using \textit{Chandra} data, to the one obtained by the \textit{Planck} collaboration, using \textit{XMM-Newton} data.
\begin{table}[h]
  \caption{Scaling relation parameters for both calibration samples.}             
  \label{table:scaling_relation}      
  \centering                          
  \begin{tabular}{c c c}        
  \hline\hline                 
  X-ray sample & \textit{Chandra} & \textit{XMM-Newton} \\    
  \hline                        
  $Y^*$ & $-0.29\pm0.01$ & $-0.19\pm0.02$ \\
  $\alpha$ & $1.70\pm0.1$ & $1.79\pm0.08$ \\
  $(1-b)$ & $0.89 \pm 0.04$ & $0.78 \pm 0.04$ \\
  scatter & $21\%$ & $18\%$ \\
     
  \hline                                   
  \end{tabular}
  \tablefoot{In the case of the \textit{XMM-Newton} sample, only the bias was re-estimated in this work, as a full re-calibration of the scaling relation did not 
  yield different results to \cite{planck_collaboration_planck_2014}.}
\end{table}
The scaling relations differ in both normalisation and slope. The lower normalisation obtained in this work is expected given the fact that cluster temperatures measured by \textit{Chandra} are systematically higher 
than those measured by \textit{XMM-Newton}, leading to higher masses \citep[see][]{schellenberger_xmm-newton_2015, potter_hydrostatic_2023}. Taking the non-unitary slope of the relation into account, 
the difference of 20\% in normalisation corresponds to a difference in mass of 16\%, which corresponds exactly to the expected difference in mass of 16\% calculated 
from the temperature discrepancy between the two telescopes in \cite{schellenberger_xmm-newton_2015}. It is important to note that the change in normalisation is degenerate with the 
change in hydrostatic mass bias. Thus, in the fitting process, normalisation and mass bias fully compensate each other given the use of the weak-lensing data as a 'true' mass anchor.\\

  \begin{figure*}[]
    \centering
    \resizebox{\hsize}{!}
          {\includegraphics{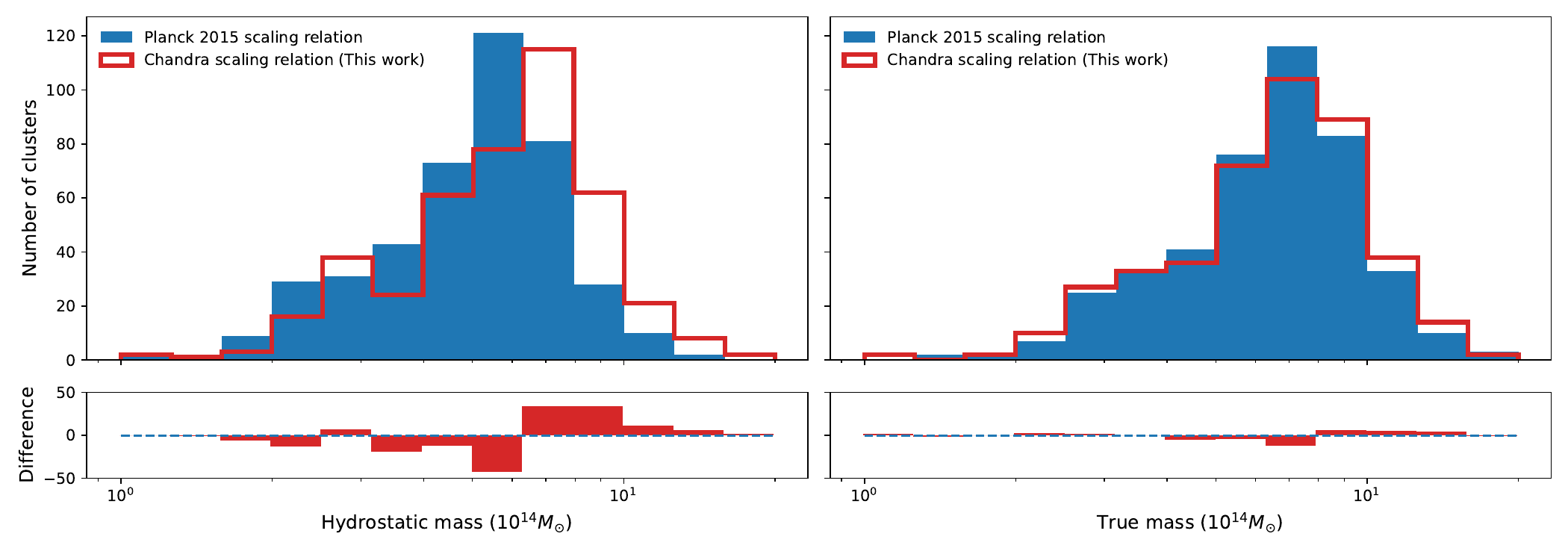}}
    \caption{Mass distributions of the clusters of the full \textit{Planck} cosmological sample, when using the \textit{Planck} 2015 scaling relation in blue and the relation calibrated in this work in red. The distributions in both panels are obtained from the same sample and identical SZ data.\\
    \textbf{Left:} Hydrostatic masses i.e. not including a $(1-b)$ correcting factor, as obtained from Eq. \ref{eq:M_500_SZ}.\\
    \textbf{Right:} 'True' masses, obtained from the full scaling relations including the bias corrections presented in Table \ref{table:bias}: $M_\text{true}=\frac{M_\text{hydrostatic}}{(1-b)}$.}
        \label{fig:M_dist_cosmo_sample}
  \end{figure*}

The shallower slope can also be explained by the instrumental differences, as the difference in temperature and therefore mass is larger at larger temperatures and masses. 
This leads to the difference of mass between the least and most massive clusters to be greater in the \textit{Chandra} sample than in the \textit{XMM-Newton} sample, and therefore to a shallower slope. We note that the slope 
obtained in this work is closer to the expected self-similar value of $5/3$ \citep{kravtsov_new_2006}.\\
Figure~\ref{fig:M_dist_cosmo_sample} shows the mass distribution of the clusters for the full \textit{Planck} cosmological sample 
when using the \textit{Planck} 2015 scaling relation in blue and the relation calibrated in this work in red. In the left panel, the masses are hydrostatic masses , that is, not including a $(1-b)$ correcting factor, 
while in the right panel, they are 'true' masses, obtained from the full scaling relations, including the bias corrections presented in Table \ref{table:bias}.
Before bias correction, the change of normalisation of the scaling relation due to differences in instruments and possibly samples is reflected in a shift of the distribution peak, with the \textit{Planck} scaling relation 
leading to a peak at lower masses. However, the use of the CCCP/MENeaCS sample as a common 'true' mass calibration source and the complete degeneracy between the normalisation and the bias in the scaling relation 
lead to final mass distributions that peak at the same mass regardless of the calibration sample used. On the other hand, the change of slope in the scaling relation is not degenerate with the bias and
does have a slight effect on the mass distribution, leading to a less peaked distribution, with more lower and higher mass clusters.\\

In terms of uncertainties and scatter, the scaling relations are similar. While this is expected for scatter, one would expect that a scaling relation calibrated on a larger sample would have smaller 
uncertainties. While this holds for the $Y_{\text{SZ}}-M_{500}^{Y_{\text{X}}}$ relations, it is no longer true for the final $Y_{\text{SZ}}-M_{500}$ relations, for which the uncertainty on the slope is larger in this work. 
This can be explained by the difference in the $M_{500}^{Y_{\text{X}}}-M_{500}$ relation, which is different for the \textit{XMM-Newton} and \textit{Chandra} samples since the X-ray observations were obtained from different instruments. 
We used Eq. \ref{eq:Y_X-M_500} to derive $M_{500}^{Y_{\text{X}}}$ 
for the \textit{Chandra} sample, while the \textit{Planck} collaboration used the $Y_{\text{X}}-M_{500}$ relation from \cite{arnaud_universal_2010} for the \textit{XMM-Newton} sample. The latter has lower uncertainties, which when propagated, 
lead to similar uncertainties in the final $Y_{\text{SZ}}-M_{500}$ relation, even though the sample used in this work is larger.
\section{Constraining cosmological parameters}
\label{cosmo}
To constrain the cosmological parameters $\Omega_{\text{m}}$ and $\sigma_8$, we use the procedure from \cite{planck_collaboration_planck_2016}, implemented in the \texttt{CosmoMC} Fortran library \citep{lewis_cosmological_2002}. 
This section briefly summarises the main steps of the procedure, and the reader is referred to \cite{planck_collaboration_planck_2016} for more details.
\subsection{Modelling and likelihood}
\subsubsection{Number counts modelling}
\label{modelling}
To constrain the cosmological parameters $\Omega_{\text{m}}$ and $\sigma_8$ using cluster number counts, we need a theoretical prediction of cluster abundance as a function of cosmology. We use the mass function from \cite{tinker_toward_2008}, 
which predicts the number of halos per unit mass and volume. On the observation side, the 439 clusters in the \textit{Planck} cosmological sample are detected using the MMF3 algorithm, based on their SZ signal-to-noise ratio, 
with a detection threshold of $S/N=6$ to ensure a high purity of the sample. 
Follow-up observations allowed for redshift determinations. We can thus model the observed cluster number counts as a function of signal-to-noise and redshift, and compare this to the theoretical predictions 
using the following equation:\\
\begin{equation}
  \label{eq:number_counts}
  \frac{\text{d}N}{\text{d}z \text{d}q}=\int \text{d}\Omega_{\text{mask}} \int \text{d}M_{500} \frac{\text{d}N}{\text{d}z \text{d}M_{500} \text{d}\Omega} P[q\mid\bar{q}_{\text{m}}(M_{500},z,l,b)],
\end{equation}
where $\frac{\text{d}N}{\text{d}z \text{d}q}$ is the observed number counts, 
$\frac{\text{d}N}{\text{d}z \text{d}M_{500} \text{d}\Omega}$ is the product of the mass function times the volume element, 
$q$ is the signal-to-noise ratio, and $\bar{q}_{\text{m}}(M_{500},z,l,b)$ is the mean signal-to-noise ratio of a 
cluster of mass $M_{500}$ at redshift $z$. This last term is where the scaling relation, calibrated in the previous section, is used to relate the mass and redshift of a cluster to its expected signal-to-noise ratio:\\
\begin{equation}
  \label{eq:q_m}
  \bar{q}_{\text{m}}=\bar{Y}_{SZ}(M_{500},z) / \sigma_{\text{f}}[\bar{\theta}_{500}(M_{500},z),l,b],
\end{equation}
where $\bar{Y}_{SZ}(M_{500},z)$ is the mean SZ signal of a cluster of mass $M_{500}$ at redshift $z$ obtained from the scaling relation, 
and $\sigma_{\text{f}}$ is the noise of the detection filter at the cluster position $(l,b)$ and aperture $\bar{\theta}_{500}$. Given a certain cosmology, the aperture is directly linked to the mass and redshift:\\
\begin{equation}
  \label{eq:theta_500}
  \bar{\theta}_{500}=\theta_{*} \left[\frac{h}{0.7}\right]^{-2/3} \left[\frac{(1-b)M_{500}}{3.10^{14}M_{\odot}}\right]^{1/3} E^{-2/3}(z) \left[\frac{D_A(z)}{500 \, \text{Mpc}}\right]^{-1},
\end{equation}
with $\theta_* = 6.997 \text{ arcmin}$. The uncertainties and intrinsic scatter of the scaling relation, as well as the noise fluctuations and selection function of the survey are accounted for by the distribution $P[q\mid\bar{q}_{\text{m}}]$.
Following \cite{planck_collaboration_planck_2016}, we use the analytical approximation of the selection function of the \textit{Planck} survey that assumes pure Gaussian noise, leading to an error function form 
for the selection function.
\subsubsection{Likelihood}
\label{likelihood}
As in \cite{planck_collaboration_planck_2016}, we use a likelihood constructed in the 2D space of redshift and signal-to-noise ratio, dividing it into 10 redshift bins of width $\Delta z=0.1$ and five signal-to-noise 
bins of width $\Delta \text{log}q=0.25$. Modelling the observed cluster number counts $N(z_i,q_i)=N_{ij}$ in each bin as independent Poisson random variables with mean rates $\bar{N}_{ij}$, the log-likelihood 
is the following:\\
\begin{equation}
  \label{eq:likelihood_cosmo}
  \text{ln}\mathcal{L}=\sum_{i,j}\left[N_{ij} \text{ln}\bar{N}_{ij}-\bar{N}_{ij}-\text{ln}\left(N_{ij}!\right)\right].
\end{equation}
Equation~\ref{eq:number_counts} is used to obtain the theoretically predicted mean rates $\bar{N}_{ij}$ according to the cosmological parameters and scaling relation:\\
\begin{equation}
  \label{eq:mean_rates}
  \bar{N}_{ij}= \frac{\text{d}N}{\text{d}z \text{d}q}(z_i,q_j) \Delta z \Delta q_j.
\end{equation}
We map the parameter space using the \texttt{CosmoMC} Monte-Carlo-Markov-Chain sampler using the likelihood in Eq.\ref{eq:likelihood_cosmo} to constrain the cosmological parameters and obtain the confidence intervals. 
We consider the standard $\Lambda$CDM model, varying the 6 parameters: the baryon density $\Omega_b h^2$, the cold dark matter density $\Omega_c h^2$, the angular size of the sound horizon at recombination $\theta_s$, 
the optical depth to reionisation $\tau$, the amplitude of the primordial power spectrum $A_s$, and the spectral index $n_s$. The parameters of the scaling relation are also sampled within the Gaussian priors 
obtained in Sect. \ref{scaling_relation}.\\
Similarly to \cite{planck_collaboration_planck_2016}, we need additional information to constrain parameters that cluster number counts are not sensitive to 
(i.e. all parameters but $\Omega_{\text{m}}$ and $\sigma_8$). 
We add a prior from Big Bang nucleosynthesis \citep[from][]{steigman_neutrinos_2008} for the value of $\Omega_b h^2= 0.0218\pm0.0012$ and priors from CMB anisotropies 
\citep[from][]{planck_collaboration_planck_2014-2,planck_collaboration_planck_2016-2} for the value of $n_s=0.9624 \pm 0.014.$ and $\tau=0.055\pm0.009$. While priors from more recent studies could be used, 
we chose to keep the same priors as in \cite{planck_collaboration_planck_2016} for full consistency. It should also be noted that while these priors are necessary to generate the halo mass function, their 
precise value has a very limited impact on the final $\Omega_{\text{m}}$ and $\sigma_8$ constraints. We also use the Baryon Acoustic Oscillation (BAO) data 
from \cite{alam_clustering_2017} (BOSS DR12) as an external dataset, using the standard \texttt{CosmoMC} implementation to include it in the likelihood, to help constrain the value of $H_0$ which the number counts alone are unable to. 

In the following, we report the constraints in terms of the derived parameters $\Omega_{\text{m}}$ and $\sigma_8$, obtained from the underlying parameters $\Omega_b h^2$, $\Omega_c h^2$, $\theta_s$, $\tau$, $A_s$, and $n_s$.

\subsection{Cosmological constraints}
\begin{figure}[]
    \centering
    \includegraphics[width=\columnwidth]{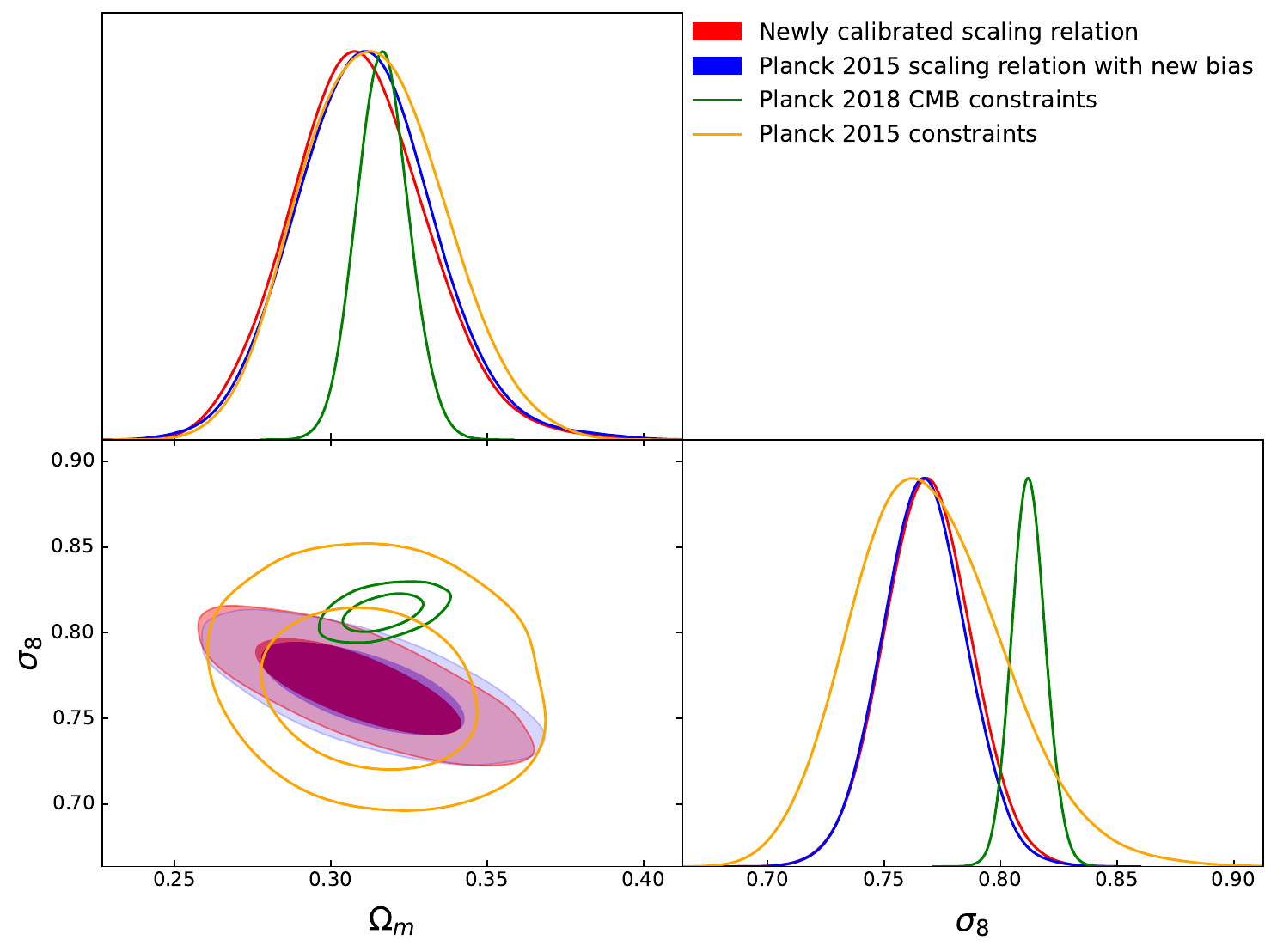}
    \caption{Final cosmological constraints obtained with the scaling relation calibrated in Sect. \ref{scaling_relation} (in red) 
    and comparison with constraints from SZ number counts obtained in \cite{planck_collaboration_planck_2016} (in yellow), 
    constraints from CMB primary anisotropies from \cite{planck_collaboration_planck_2020} (in green), and constraints obtained using the scaling 
    relation from \cite{planck_collaboration_planck_2016} with the bias calibrated in Sect. \ref{bias} (in blue).}
    \label{fig:constraints}
\end{figure}

Figure \ref{fig:constraints} shows the cosmological constraints obtained in the $\Omega_m - \sigma_8$ plane, and compares them to the ones obtained in \cite{planck_collaboration_planck_2016}, the latest constraints 
from the CMB anisotropies observed by \textit{Planck} \citep{planck_collaboration_planck_2020}, and the constraints obtained when using the scaling relation from \cite{planck_collaboration_planck_2016} with the bias 
calibrated in this work. Like all other triangle plots in this work, the \texttt{GetDist} Python package \citep{lewis_getdist_2019} was used to obtain contours from the MCMC chains.

Table \ref{table:cosmo} shows the constraints on the cosmological parameters for the scaling relation calibrated in Sect. \ref{scaling_relation} and the \textit{Planck} scaling relation, updated 
with the bias calibrated in Sect. \ref{bias} (\textit{XMM-Newton} D+nD case) as it is the fairest comparison point to understand the effect of changing only the X-ray calibration sample.
\begin{table}
  \caption{Constraints on $\Omega_{\text{m}}$, $\sigma_8$, and $S_8 \equiv \sigma_8 \sqrt{\Omega_m / 0.3}$ obtained with the scaling relations presented in Table \ref{table:scaling_relation}.}             
  \label{table:cosmo}      
  \centering                          
  \begin{tabular}{c c c}        
  \hline\hline                 
  X-ray sample & \textit{Chandra} & \textit{XMM-Newton} \\    
  \hline                        
  $\Omega_{\text{m}}$ & $0.309\pm0.022$ & $0.311\pm0.022$ \\
  $\sigma_8$ & $0.769\pm0.019$ & $0.767\pm0.018$  \\
  $S_8$ & $0.780\pm0.019$ & $0.781\pm0.020$ \\
     
  \hline                                   
  \end{tabular}
\end{table}
The constraints obtained with either calibration sample and the biases computed in this work are in full agreement, both in terms of the central value as well as uncertainties. This is quite remarkable, given the fact that the two samples differ not only in terms of selection function, with the \textit{Chandra} sample being SZ-selected and the \textit{XMM-Newton} sample assembled from pre-existing X-ray observations, but also in terms of
instruments, with the two samples coming from telescopes known to yield different results in terms of temperature and thus mass estimates, when observing the same clusters 
\citep{schellenberger_xmm-newton_2015, potter_hydrostatic_2023}.\\
Additionally, the constraints obtained in this work are in agreement with the ones originally obtained in \cite{planck_collaboration_planck_2016}, with similar central values albeit better constraining power 
in the $S_8$ direction, due to an improved calibration of the mass bias, thanks to the larger weak-lensing dataset from \cite{herbonnet_cccp_2020} used in this re-analysis.\\
\section{Discussion}
\label{discussion}
\subsection{Possible sources of systematic uncertainties}
\label{systematics}

In this Section, we provide an overview of several sources of systematic error that could cause a shift in the cosmological constraints, and try to quantify the effect whenever possible.

\paragraph{Mass function}
Like every study of galaxy cluster number counts, a choice of reference halo mass function has to be made when comparing the observed cluster abundance with the theoretical predictions. For our analysis, we chose to use the \cite{tinker_toward_2008} mass function for consistency with the baseline \cite{planck_collaboration_planck_2016} analysis. In the Planck analysis, changing the halo mass function was found to shift the final value of $\Omega_{\text{m}}$ up by $\sim$10$\%$ and the final value of $\sigma_8$ down by $\sim$10$\%$ as well. Since we use the same cluster catalogue and simply change the mass calibration, we expect a very similar shift in the final constraints if we were to change the mass function to that of \cite{watson_halo_2013} as well. More recently, \cite{abbott_dark_2020} and \cite{bocquet_spt_2024} have tried to quantify the uncertainties around the \cite{tinker_toward_2008} mass function using other simulations. \cite{bocquet_spt_2024} compared marginalising over those uncertainties with not accounting for them and found no significant shifts in the final constraints, thus concluding that the uncertainties on the halo mass function could be neglected.

\paragraph{Weak-lensing mass bias}
Calibrating the mass bias from weak-lensing data requires the assumption that the weak-lensing mass estimates are unbiased. 
However, it has been shown that weak-lensing mass estimates can be subject to systematic biases \citep[see e.g.][]{grandis_srgerosita_2024, bocquet_spt_2023}. 
In this work, we do not explicitly correct for these systematic biases as we use masses taken from \cite{herbonnet_cccp_2020} which are already 
corrected by comparing with simulations \citep[see Appendix C of][for an extensive description of the correction procedure]{herbonnet_cccp_2020}.

\paragraph{Mass calibration at fixed cosmology}
A caveat of the approach is that the scaling relation is obtained for a fixed cosmology, while the value of the parameters varies during the MCMC fitting. The reason for this is two-fold: 
first, since this work aims at studying the impact of the mass calibration, we thus chose to keep all other elements of the procedure identical to \cite{planck_collaboration_planck_2016}. 
Also, a fully consistent approach would require the re-extraction of X-ray data at every step, as a fixed cosmology is also assumed during the data reduction process. Overall, the effect of varying the cosmology within the $2 \sigma$ contours 
we obtain yields a maximum difference of $2 \%$ in X-ray mass for the highest redshift clusters in the calibration sample, and less than $1 \%$ for most of the sample. This is smaller than the uncertainties on the mass calibration, and we thus expect the effect to be negligible with respect to the other sources of uncertainty.

\paragraph{Miscentering of MMF detections} 
When calibrating the scaling relation, we have access to a very precise position and size prior from the X-ray data, as the \textit{Chandra} telescope has a very good angular resolution of 0.5 arc-seconds (6 arc-seconds for \textit{XMM-Newton}). We make use of that prior to perform pointed detection with the MMF algorithm in the Planck data, as blind detections are not only more computationally expensive, but also less precise due to \textit{Planck}'s larger point spread function of $\sim$7 arc-minutes. However, the PSZ2 catalogue that we use to constrain the cosmology is obtained from blind detections of clusters, which could cause biases in the retrieved SZ signal. This issue was investigated in \cite{planck_collaboration_planck_2014}, where a good agreement was found between the signal-to-noise ratio for blind and pointed detections (see Fig. 2 of that paper). Given the fact that we are binning the clusters into quite large S/N bins, we expect the small scatter between the two S/N estimates to be negligible.

\paragraph{Correct retrieval of the scaling relation} 
With the increased precision of the mass calibration via scaling relation presented in this work, we need to ensure that the accuracy of the method is sufficient to not induce a systematic bias in the final constraints. To verify this, we generated 100 synthetic cluster samples resembling the \textit{Chandra} sample with realistic observables, following a known underlying scaling relation (see Appendix \ref{toy_model} for details on the sample generation). We then ran these samples through the scaling relation calibration pipeline, and compared the retrieved relations with the true underlying relation.

\begin{figure}[]
  \centering
  \includegraphics[width=\columnwidth]{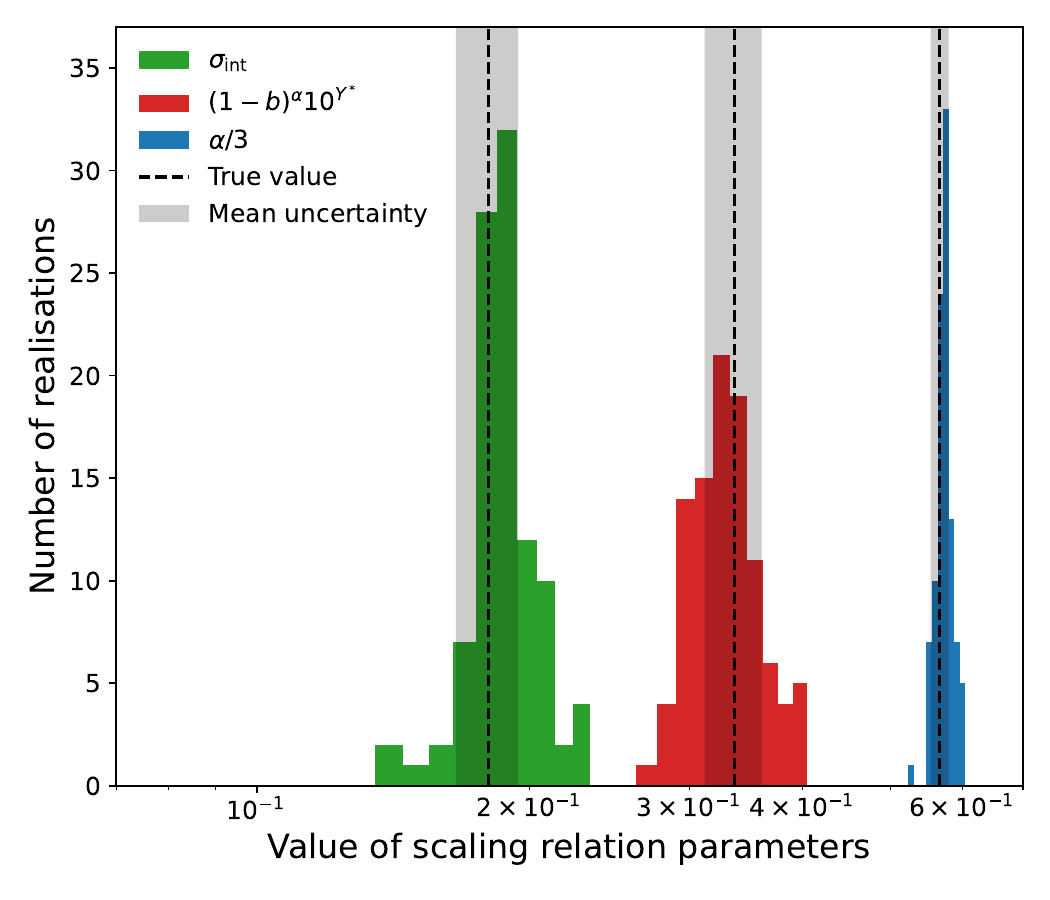}
  \caption{Retrieved $Y_{\text{SZ}}-M_{500}$ scaling relation on 100 synthetic calibration samples compared with the true underlying relation used for sample generation. The retrieved slope (divided by three for better readability) is shown in blue, the retrieved intrinsic scatter in green, and the relevant combination of the degenerated parameters $(1-b)$ and $Y^*$ in red. The true value of each parameter is shown by the black dashed lines and the mean uncertainty returned by the calibration procedure is shown by the grey contours.}
    \label{fig:toy_model_res}
  \end{figure}
The results of the comparison are presented in Fig. \ref{fig:toy_model_res}, which shows that the calibration procedure retrieves the underlying scaling relation with very good accuracy and correctly estimated uncertainties. In particular, the parameter combination $(1-b)^{\alpha}10^{Y^*}$, which sets the mass scale and is by far the most important parameter in terms of final cosmological constraints is perfectly retrieved. In comparison, the very slight bias of around 0.5$\sigma$ (the mean retrieved value being 1.72 for a true value of 1.70) that can arguably be observed on the slope would result in no shift at all in final cosmological constraints. Indeed, identical constraints are obtained with the \textit{Chandra} and \textit{XMM-Newton} calibration sample, even though they result in a slope difference four times larger than the bias found here. Similarly, the difference between the mean retrieved intrinsic scatter and the true value is much smaller than the difference between those obtained with the two calibration samples, and would thus not affect the final constraints at all. Using the same population generation procedure, we also show that the possible presence of correlated intrinsic scatter between $Y_\text{SZ}$ and $Y_{\text{X}}$ would not lead to a significant bias in the final constraints (see Appendix \ref{toy_model} for more details).

\paragraph{Double systems}
Given the $\sim$7 arc-minutes Planck beam, two clusters that are roughly aligned along the line of sight can lead to a single detection in the Planck maps. During the mass calibration process, we can avoid this problem thanks to the much higher resolution X-ray images that allow us to remove such double systems from the calibration sample. Nevertheless, these systems cannot be identified when compiling the PSZ2 catalogue used to constrain the cosmology. Given the abundance of double systems in the calibration, we can expect around 16 out of the 439 clusters in the PSZ2 sample to be double systems. It is hard to predict the effect of the presence of double systems on the final results as it is not possible to know a priori the effect on the S/N of detection. We can however expect that the magnitude of the bias in the final constraints is small given the small number of double systems that are likely present in the PSZ2 sample.

\paragraph{Implementation of the likelihood}
The implementation of the likelihood is the same as in \cite{planck_collaboration_planck_2016}, which was validated by comparing three independent implementations which were found to yield differences in cosmological constraints of less than 1/10 $\sigma$. Given that the maximal improvement of constraining power compared to \cite{planck_collaboration_planck_2016} is a factor of two on the value of $S_8$, the uncertainty coming from the implementation of the likelihood is still negligible in this work.
\\
\\
In conclusion, the halo mass function is the main source of systematic uncertainty. The magnitude of this systematic uncertainty is such that all other effects are negligible in comparison. The calibration of the halo mass function is the subject of ongoing research and debate in the community, and as such is outside the scope of this study. Nevertheless, the recent results of \cite{bocquet_spt_2024} seem to indicate that the impact of the uncertainty on the mass function on SZ cluster number count cosmology studies might not be as important as that estimated in \cite{planck_collaboration_planck_2016}, with recent simulation results being in reasonably good agreement with each other.

\subsection{Exploring the redshift dependence}
\label{redshift_dependance}
\begin{figure}[]
  \centering
  \includegraphics[width=\columnwidth]{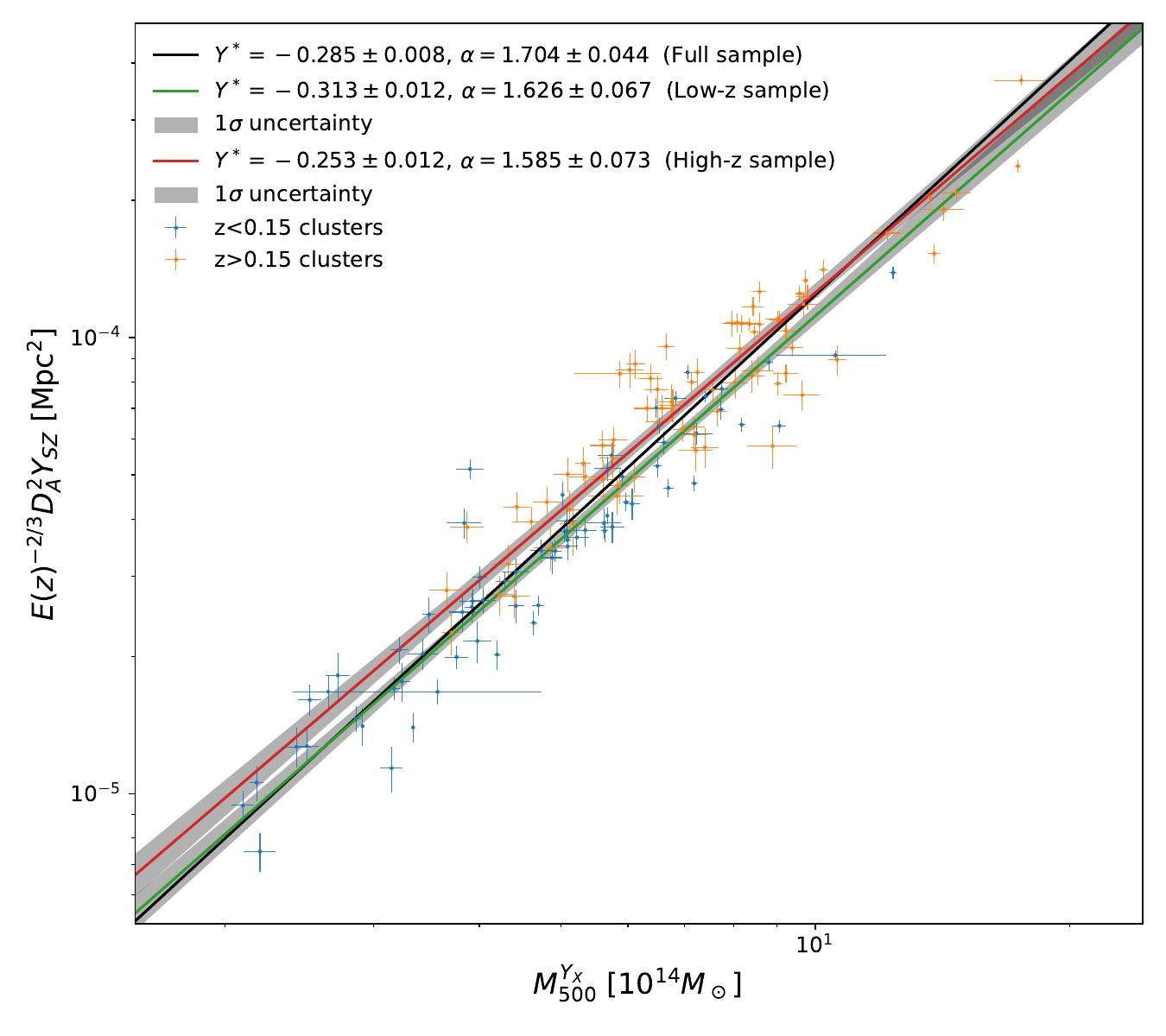}
  \caption{Calibration of the $Y_{\text{SZ}}-M_{500}^{Y_{\text{X}}}$ scaling relation for the high (best fit in red) and low (best fit in green) redshift subsamples. 
  The black line shows the best-fit relation obtained previously on the full sample, and the grey contours correspond to $1\sigma$ uncertainties for each subsample fit.}
    \label{fig:scaling_z_subsamples}
  \end{figure}

\begin{figure}[]
  \centering
  \includegraphics[width=\columnwidth]{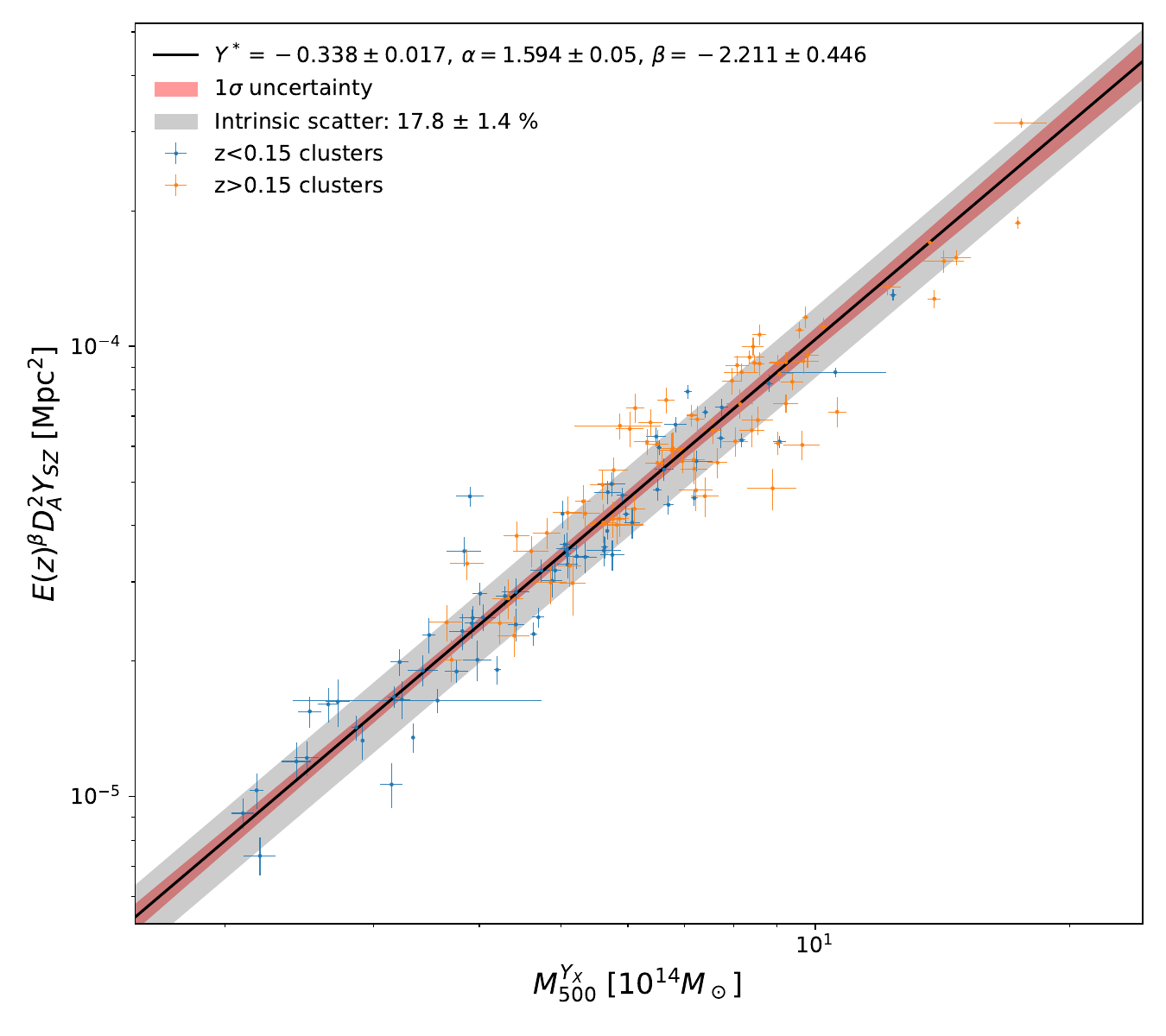}
  \caption{Calibration of the $Y_{\text{SZ}}-M_{500}^{Y_{\text{X}}}$ scaling relation, letting the redshift dependence vary as a free parameter. 
  The high and low redshift sub-samples are shown in different colours to help visualise the effect of 
  the free redshift dependence, but the fit is performed on the whole sample at once. The black line shows the best-fit relation, with red contours corresponding to $1\sigma$ uncertainties, 
  and grey contours to the intrinsic scatter.}
    \label{fig:scaling_free_z_dependance}
  \end{figure}

\begin{figure}[]
  \centering
  \includegraphics[width=\columnwidth]{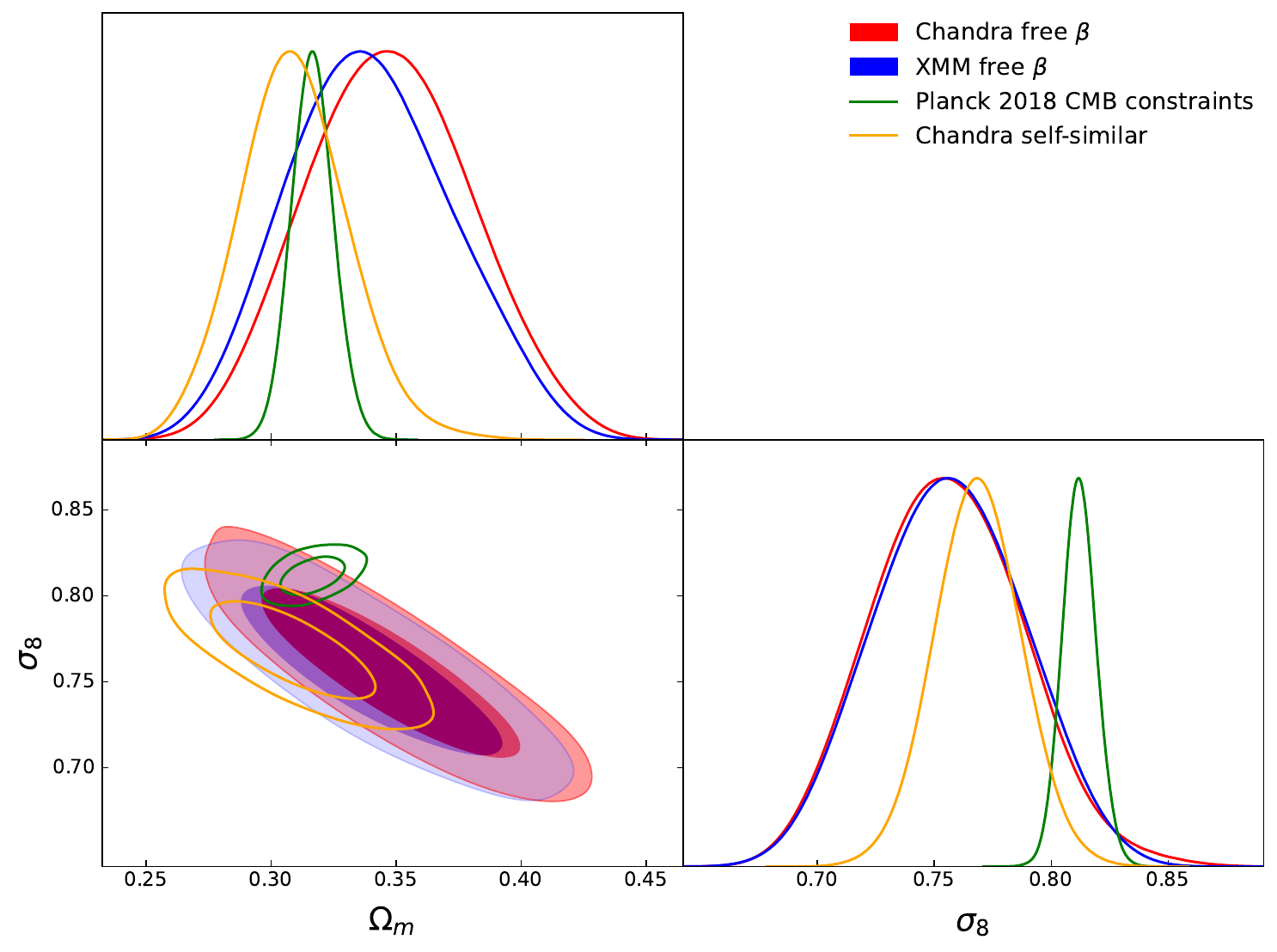}
  \caption{Cosmological constraints obtained with the redshift dependence fitted from the data, with both \textit{Chandra} (in red) and \textit{XMM-Newton} (in blue) calibration samples, 
  and comparison with the results obtained with the fixed, self-similar redshift dependence using the \textit{Chandra} calibration sample (in yellow) and \cite{planck_collaboration_planck_2020} 
  constraints from CMB primary anisotropies (in green).}
  \label{fig:constraints_free_z_dependance}
\end{figure}

When calibrating the scaling relation in Sect. \ref{scaling_relation}, we fixed the redshift dependence of the relation to the self-similar value of $-2/3$ in order to follow the calibration process of 
\cite{planck_collaboration_planck_2014}. However, if we split the sample into two sub-samples with clusters below or above the median redshift of the full sample of $z=0.15$ and calibrate the scaling relation 
on each sub-sample independently, the resulting best fits are in disagreement. Fig. \ref{fig:scaling_z_subsamples} shows the scaling relations calibrated on the two sub-samples, and compares them with the best fit 
obtained on the full sample.
For both sub-samples, the best-fit relation has a shallower slope than the one obtained on the full sample by more than $1\sigma$, and the low-z (respectively high-z) best-fit relation has a lower (resp. higher) 
normalisation than the scaling relation obtained with the full sample. This suggests that the assumed self-similar redshift dependence cannot properly fit the data over the entire redshift range of the sample.\\
To explore this further, we fit the scaling relation with a free redshift dependence, using the following relation:
\begin{equation}
  \label{eq:Y_SZ-M_500_free_z}
  E^{\beta}(z)\frac{D^2_A\,Y_{\text{SZ}}}{Y_{\text{piv}}}=10^{Y^*}\left(\frac{M_{500}^{Y_{\text{X}}}}{M_{\text{piv}}}\right)^{\alpha}.
\end{equation}
We add an additional free parameter to the equation, letting the power of $E(z)$ vary from the self-similar value of $-2/3$, and modify the likelihood presented in Sect. \ref{calibration_scaling_relation} to 
account for this additional degree of freedom. We use a broad, flat prior on $\beta$. Fig. \ref{fig:scaling_free_z_dependance} shows the best-fit relation obtained with this model.
The best-fit value for the redshift dependence is $\beta=-2.22\pm0.45$, which is in tension with the self-similar value of $-2/3$ at more than 3$\sigma$. Introducing this degree of freedom in the scaling relation also 
reduces the intrinsic scatter ($\sigma_{int}=17.8\%$ instead of $19.6\%$), and lowers the mass dependence to $\alpha=1.59 \pm 0.05$, below the $5/3$ value predicted by the self-similar model by 1$\sigma$.\\
To verify if this is simply a trend in this particular sample, we repeat the same analysis on the \textit{XMM-Newton} sample. We obtain a best-fit value of $\beta=-1.96\pm0.47$, in agreement with the one 
obtained on the \textit{Chandra} sample and in roughly 3$\sigma$ tension with the self-similar value. We also observe a similar decrease in intrinsic scatter ($14.8\%$ from $15.8\%$), and the mass dependence that initially was steeper than the 
self-similar value is in perfect agreement with it at $1.66\pm0.078$.\\
\cite{andreon_important_2014} performed a similar calibration of the Planck $Y_{\text{SZ}}-M_{500}$ relation letting the redshift dependence vary, and found $\beta=-2.5\pm0.4$, which is in agreement with the results found here. Another result we can compare with is the redshift-dependent bias presented in Sect. 6.4 of \cite{sereno_comalit_2017}. The authors computed the best-fit bias between the weak-lensing masses they compiled and the Planck masses (obtained with the scaling relation from \cite{planck_collaboration_planck_2014}, i.e. with a self-similar redshift dependence) for a set of clusters where both mass estimates were available. They allowed for both redshift- and mass-dependence, and found
\begin{equation}
  \label{eq:M_SZ-M_WL_Comalit}
  M_{\text{SZ}}=\left(1.05\pm0.11\right) M_{\text{WLc}}^{0.94\pm0.08} \left(\frac{E(z)}{E(z_{\text{ref}})} \right)^{-2.40\pm0.63}.
\end{equation}
While the deviation of the mass dependence from unity is negligible, their result points to a strong redshift dependence, with SZ-masses of high-redshift clusters being biased lower with respect to WL-masses than those of low-redshift clusters. If we were to recompute this bias using SZ-masses obtained with the modified scaling relation presented in Table \ref{table:scaling_free_z}, we would expect the redshift dependence of the bias to change, as it would be affected by the non-self-similar redshift evolution of the scaling relation used to compute the SZ-masses. While performing the fit is outside the scope of this study, we can expect a change of redshift dependence of the order of $\frac{-2/3 - \beta}{\alpha} \sim 1$, raising the best-fit value of the redshift dependence from $-2.4$ to roughly $-1.4$. While an even larger departure from the self-similar redshift evolution of the $Y_{\text{SZ}}-M_{500}$ relation would be needed to fully explain the redshift dependence of the bias, the deviation from self-similar redshift evolution computed in this study can explain a significant part of the effect observed in \cite{sereno_comalit_2017}. Other sources of biases, in the weak-lensing mass computation for example, are still left to be explored and could explain the rest of the redshift dependence of the SZ-mass to WL-mass bias.\\ 
\cite{wicker_constraining_2023} investigated the evolution of the hydrostatic mass bias with mass and redshift by studying the evolution of the gas fraction in galaxy clusters. They found
\begin{equation}
  \label{eq:bias_wicker}
  (1-b)=\left( 0.83 \pm 0.04 \right) \left( \frac{M}{\langle M \rangle} \right) ^{-0.06 \pm 0.04} \left(\frac{1+z}{\langle 1+z \rangle} \right)^{-0.64 \pm 0.18}.
\end{equation}
Similarly to \cite{sereno_comalit_2017}, the mass dependence is negligible, but they found a significant negative redshift dependence. Following the same reasoning as above, the deviation from self-similar redshift evolution found in this work would correspond to redshift evolution of $(1-b)$ of around $-1$, which, while not fully compatible, is coherent with the negative value found by \cite{wicker_constraining_2023}.\\
Given the fact that both calibration samples show a similar and significant deviation from the self-similar redshift dependence, even though their selection function is very different (SZ-selected for \textit{Chandra}, 
pre-existing X-ray observation for \textit{XMM-Newton}), and that other results in the literature point to a similar deviation, we conclude that this is likely not a selection effect, but rather a systematic effect, stemming either from actual physical processes not properly 
described by the self-similar model or from untreated systematics in the instrument or data extraction.\\
\cite{lovisari_x-ray_2017} provides morphological information for 92 clusters in the \textit{Chandra} sample, which we can make use of to further investigate this redshift dependence question. 
When fitting the redshift dependence of the scaling relation on the 32 clusters with a relaxed morphology, the 
best fit value is $\beta = -2.5\pm1.3$, which is steeper than the value obtained on the full sample (while still being in general agreement). 
When fitting the redshift dependence of the scaling relation on the 60 clusters with a mixed or disturbed morphology, the 
best-fit value is $\beta = -1.1\pm0.8$, which is shallower than the value obtained on the full sample, and closer to the self-similar value of $-2/3$. Although the statistical significance is low, 
this suggests that the deviation from the self-similar model might be related to the dynamical state of the clusters, and that relaxed, 
high redshift clusters exhibit a lack of $Y_{\text{SZ}}$ signal compared to their X-ray derived mass. A possible explanation 
for this behaviour is the fact that \textit{Planck}-detected clusters at redshifts above $z$$\sim$$0.15$ are smaller than the beam of the instrument. This leads to a large smoothing of the cluster signal, 
which is accounted for in the MMF detection by using templates of both the cluster profile and the beam. Since neither of these templates is guaranteed to be fully accurate, this could lead to an incorrect 
extraction of the $Y_{\text{SZ}}$ signal, especially for relaxed clusters, which likely have a more peaked profile, prone to larger smoothing.\\
We use these newly obtained scaling relations to constrain cosmological parameters, and compare the results with the ones obtained with the self-similar scaling relations. We follow the procedure described in Sect. \ref{scaling_relation} and \ref{cosmo}, 
with the only difference being the use of the new scaling relations. The first step is to combine these new $Y_{\text{SZ}}-M_{500}^{Y_{\text{X}}}$ relations with the respective $M_{500}^{Y_{\text{X}}}-M_{500}$ relations 
to obtain the $Y_{\text{SZ}}-M_{500}$ relations. From these scaling relations, we can extract the bias, following the same procedure as in Sect. \ref{bias}. 
Table \ref{table:scaling_free_z} shows the $Y_{\text{SZ}}-M_{500}$ relations from which cosmological constraints will be derived. 
\begin{table}
  \caption{$Y_{\text{SZ}}-M_{500}$ scaling relations with free redshift dependence for both calibration samples.}
  \label{table:scaling_free_z}      
  \centering                          
  \begin{tabular}{c c c}        
  \hline\hline                 
  X-ray sample & \textit{Chandra} & \textit{XMM-Newton} \\     
  \hline                        
  $Y^*$ & $-0.34\pm0.02$ & $-0.24\pm0.03$ \\
  $\alpha$ & $1.59\pm0.1$ & $1.66\pm0.1$ \\
  $\beta$ & $-2.22\pm0.45$ & $-1.96\pm0.47$ \\
  $(1-b)$ & $0.84 \pm 0.04$ & $0.74 \pm 0.04$ \\
  scatter & $20\%$ & $17\%$ \\
     
  \hline                                   
  \end{tabular}
\end{table}
With the scaling relations fully calibrated, we can constrain the cosmological parameters, following the same procedure as in Sect. \ref{cosmo}. Fig. \ref{fig:constraints_free_z_dependance} shows the cosmological constraints obtained
with the new scaling relations, and compares them to the ones obtained with the self-similar scaling relation calibrated on the \textit{Chandra} sample. Table \ref{table:cosmo_free_z} presents the constraints on 
the cosmological parameters.
\begin{table}
  \caption{Constraints on $\Omega_{\text{m}}$, $\sigma_8$, and $S_8 \equiv \sigma_8 \sqrt{\Omega_m / 0.3}$ obtained with the scaling relations presented in Table \ref{table:scaling_free_z}.}             
  \label{table:cosmo_free_z}      
  \centering                          
  \begin{tabular}{c c c}        
  \hline\hline                 
  X-ray sample & \textit{Chandra} & \textit{XMM-Newton} \\ 
  \hline                        
  $\Omega_{\text{m}}$ & $0.347\pm0.033$ & $0.340\pm0.033$ \\
  $\sigma_8$ & $0.756\pm0.033$ & $0.757\pm0.032$  \\
  $S_8$ & $0.811\pm0.020$ & $0.803\pm0.022$ \\
     
  \hline                                   
  \end{tabular}
\end{table}
There is a significant loss in constraining power, especially along the $\Omega_m - \sigma_8$ degeneracy, given the introduction of a new degree of freedom in the scaling relation, but we can still observe a preference for 
higher $S_8 \equiv \sigma_8 \sqrt{\Omega_m / 0.3}$ values ($S_8 = 0.811 \pm 0.020$ with the \textit{Chandra} calibration sample), closer to the results of the CMB primary anisotropies of \cite{planck_collaboration_planck_2020}. 
This is due to the fact that the steeper redshift dependence, combined with the low median redshift of the X-ray and WL calibration samples ($z$$\sim$$0.2$), lead to more massive 
clusters at high redshifts than the self-similar model and thus to a higher matter density and/or more clustered matter distribution.\\
While cosmological constraints where the redshift dependence is left free to vary are presented in Fig. 6 of \cite{planck_collaboration_planck_2016}, a direct comparison is not relevant, since the fitting procedures are completely different. In this work, we fit the redshift dependence as part of the mass calibration, simultaneously fitting $Y^*$, $\alpha$, and $\beta$, then computing the hydrostatic mass bias $(1-b)$ with the previously obtained relation, and finally fitting the observed number counts to constrain the cosmology. In \cite{planck_collaboration_planck_2016}, the constraints are obtained by using the same scaling relation as in the self-similar redshift evolution case, but simply replacing the prior that fixed the redshift dependence to an uninformative flat prior when fitting the observed number counts. Since the value of the other parameters of the scaling relation is not coherently fitted with the redshift evolution, it is expected that the impact on the final constraints is not as important as in this work. In particular, no change in the final $S_8$ value is observed in \cite{planck_collaboration_planck_2016}.

\subsection{Goodness of fit}
In this analysis, there are several fitting procedures during both the mass calibration and the cosmological constraints steps for which we should investigate the goodness of fit. To evaluate the goodness of fit of the scaling relations, we present in Table \ref{table:chi2} the $\chi^2$ value compared to the number of degrees of freedom $N_{\text{DoF}}$ of the various fits performed in this work.\\
\begin{table}[h]
  \caption{$\chi^2/N_{\text{DoF}}$ value of the fits performed during the mass calibration.}             
  \label{table:chi2}      
  \centering                          
  \begin{tabular}{c c c}        
  \hline\hline                 
  X-ray sample & $Y_{\text{SZ}}-M_{500}^{Y_{\text{X}}}$ fit & $(1-b)$ fit \\    
  \hline                        
  \textit{Chandra} & $167/143$ & $75/54$ \\
  \textit{Chandra} free $\beta$ & $151/142$ & $71/54$ \\
  \textit{XMM-Newton} & $79/68$ & $77/54$ \\
  \textit{XMM-Newton} free $\beta$ & $65/67$ & $75/54$ \\
     
  \hline                                   
  \end{tabular}
\end{table}
\\
We find that all fits show $\chi^2$ values close to $N_{\text{DoF}}$, indicating a good fit of the data by the model. We also find that for both calibration samples, fitting the redshift evolution $\beta$ from the data improves the goodness of fit of the $Y_{\text{SZ}}-M_{500}^{Y_{\text{X}}}$ scaling relation as well as the hydrostatic mass bias $(1-b)$.\\
\begin{figure}[]
  \centering
  \includegraphics[width=\columnwidth]{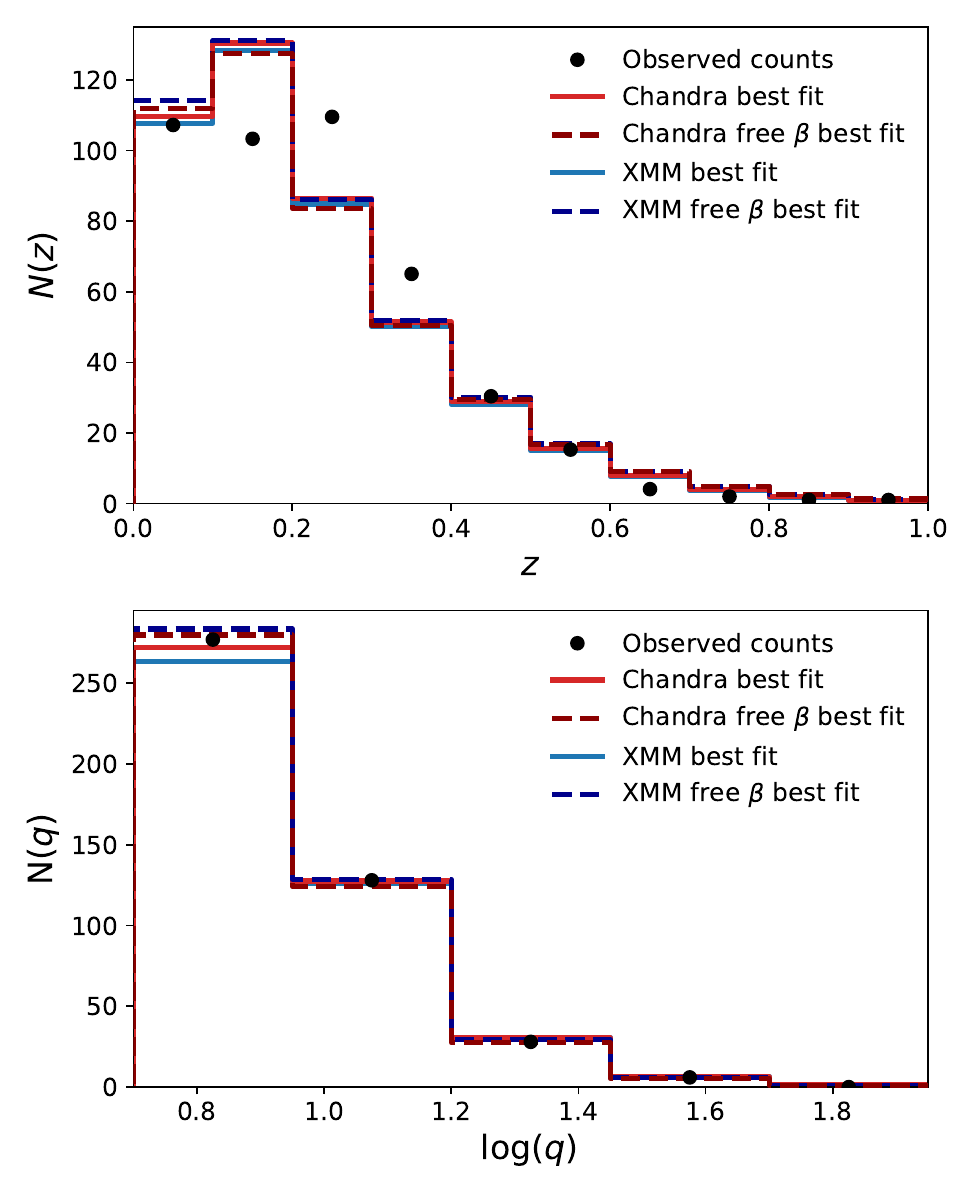}
  \caption{Redshift (top panel) and signal-to-noise (bottom panel) distribution of best-fit models from the four analysis cases presented in this work. The observed counts in the PSZ2 catalogue (q > 6) are plotted as the black points.}
  \label{fig:nc_gof}
\end{figure}
\\
Fig. \ref{fig:nc_gof} presents the redshift and signal-to-noise distribution of the best-fit models for all four analysis cases considered in this work, and compares them to the observed number counts of the PSZ2 catalogue. Since the number counts are fitted with a Poisson likelihood and some bins contain very few clusters, the most relevant statistic to evaluate the goodness of fit is the \textit{C} statistic \citep[see e.g.][for details on the \textit{C} statistic]{bonamente_distribution_2020}. Table \ref{table:cstat} presents the \textit{C} statistic value found for the four analysis cases for which cosmological parameters were fitted in this work.\\
\begin{table}[h]
  \caption{\textit{C} statistic value of the number count fits.}             
  \label{table:cstat}      
  \centering                          
  \begin{tabular}{c c}        
  \hline\hline                 
  X-ray sample & $C_{\text{best-fit}}/N_{\text{DoF}}$ \\    
  \hline                        
  \textit{Chandra} & $46.5/46$\\
  \textit{Chandra} free $\beta$ & $50/45$ \\
  \textit{XMM-Newton} & $47.1/46$ \\
  \textit{XMM-Newton} free $\beta$ & $49.3/45$ \\
     
  \hline                                   
  \end{tabular}
\end{table}
\\
We find that the fit of the number counts is acceptable but not optimal, and that the various analysis cases do not lead to a significantly different goodness of fit. This is due to the fact that the distribution of the PSZ2 clusters in redshift does not accurately follow that of the best-fit mass function prediction \citep[see also Fig.6 of][where the same problem was observed]{planck_collaboration_planck_2016}. Since this work only changes the mass calibration and has a limited impact on the best-fit cosmology, this issue is not solved.

\subsection{Comparison with other cosmological analyses}
\label{other_probes}
\begin{figure}[]
  \centering
  \includegraphics[width=\columnwidth]{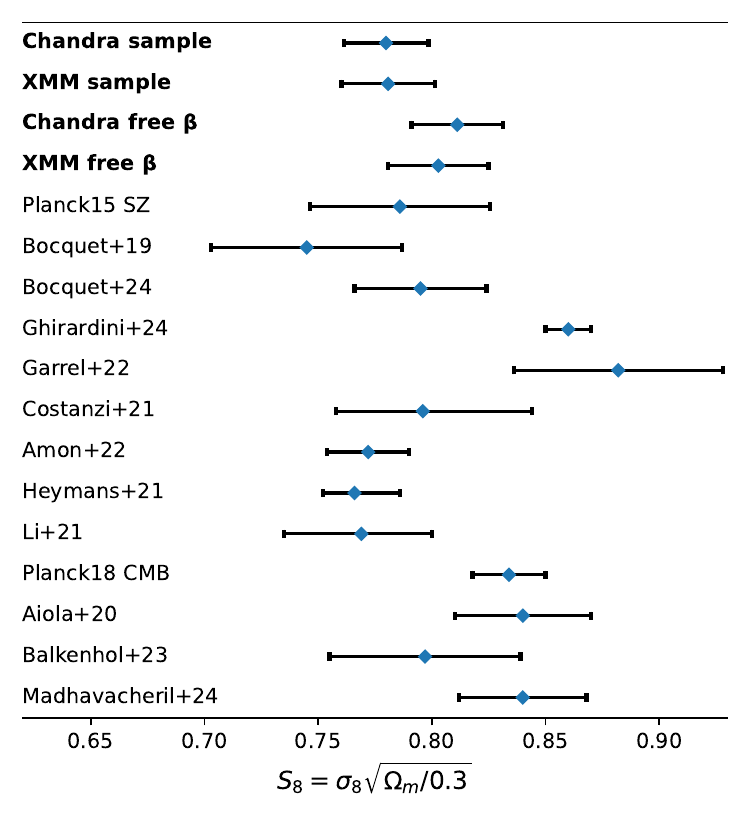}
  \caption{Comparison of the $S_8$ constraints obtained in this work (in bold) with other recent analyses.}
  \label{fig:s8_comparison}
\end{figure}

Fig. \ref{fig:s8_comparison} shows the $S_8 \equiv \sigma_8 \sqrt{\Omega_m / 0.3}$ constraints obtained in this work compared to the ones obtained in a few other recent analyses using both cluster number counts 
as well as other different cosmological probes. We selected analyses that presented constraints on $S_8$ using the definition chosen in this work for consistency. It is important to note that, while necessary to compare 
different results, this definition is not necessarily the optimal one for each probe, in the sense that errors might be smaller for another power of $\Omega_{\text{m}}$.\\
We compiled results from \cite{planck_collaboration_planck_2016} (noted Planck15 SZ), \cite{bocquet_cluster_2019} \& \cite{bocquet_spt_2024} (SPT SZ cluster number counts), \cite{ghirardini_srgerosita_2024} (eROSITA cluster number counts), \cite{garrel_xxl_2022} (analysis of an \textit{XMM-Newton}-selected cluster sample), \cite{costanzi_cosmological_2021} (DES+SPT cluster abundance),
\cite{amon_dark_2022} (DES Y3 cosmic shear), \cite{heymans_kids-1000_2021} (KiDS1000 WL + galaxy clustering), \cite{li_hyper_2023} (HSC cosmic shear), \cite{planck_collaboration_planck_2020} (noted Planck18 CMB), \cite{aiola_atacama_2020} (ACT DR4 CMB),
\cite{balkenhol_measurement_2023} (SPT-3G CMB), and \cite{madhavacheril_atacama_2024} (ACT DR6 CMB gravitational lensing). For clarity of the plot, we had to pick one HSC result amongst four different analyses, 
but \cite{dalal_hyper_2023}, \cite{miyatake_hyper_2023}, and \cite{sugiyama_hyper_2023} all find very similar results to the one chosen for the comparison.\\
\\
With this re-analysis, $S_8$ constraints from SZ number counts are competitive with the tightest constraints obtained from CMB primary anisotropies in \cite{planck_collaboration_planck_2020} 
or from the latest cosmic shear surveys in \cite{amon_dark_2022} and \cite{heymans_kids-1000_2021}, even when the redshift dependence is left free to vary. With this improved constraining power, we can use clusters to investigate the $S_8$ tension, 
as they now provide constraints tight enough to reject either the $S_8$$\sim$$0.77$ value preferred by most late-time probes or the $S_8$$\sim$$0.83$ value preferred by most CMB experiments with higher significance than 
previous results. Regardless of the calibration sample, we find that the constraints obtained with the self-similar redshift evolution agree with the values derived from the latest weak-lensing analyses 
\citep{amon_dark_2022, heymans_kids-1000_2021}, and are in tension with the \textit{Planck} CMB constraints at around 2$\sigma$. If we fit the redshift dependence of the scaling relation from the data, the derived $S_8$ constraints 
shift to higher values by about 1$\sigma$ and are right in-between values obtained by the latest WL analyses and those obtained by the latest CMB experiments, resulting in no significant tension with either result.\\

\section{Conclusion}
\label{conclusion}
Using a new full follow-up observation of the \textit{Planck} ESZ sample with the \textit{Chandra} X-ray telescope, we have calibrated a new $Y_{\text{SZ}}-M_{500}$ scaling relation using a sample of 146 clusters. Using the weak-lensing 
sample from \cite{herbonnet_cccp_2020} as an external mass calibration, we have also estimated the hydrostatic mass bias for both the newly calibrated scaling as well as the scaling relation calibrated in \cite{planck_collaboration_planck_2014}. 
When accounting for clusters not detected in the \textit{Planck} maps, we found a mass bias of $(1-b)=0.89\pm0.04$ for the new scaling relation, and $(1-b)=0.76\pm0.04$ for the \textit{Planck} scaling relation.\\
Using these scaling relations, we obtained cosmological constraints on $\Omega_{\text{m}}$ and $\sigma_8$ from the \textit{Planck} cosmological sample. We found that the constraints we obtained with both calibration samples 
are identical, with $\sigma_8 = 0.77\pm0.02$, $\Omega_m = 0.31\pm0.02$, and $S_8 \equiv \sigma_8 \sqrt{\Omega_m / 0.3}=0.78\pm0.02$. These constraints are in agreement with 
the ones obtained in \cite{planck_collaboration_planck_2016}, albeit with a much tighter constraint on $S_8$.\\
We also investigated deviations from self-similar redshift evolution of the $Y_{\text{SZ}}-M_{500}$ relations and found that letting the redshift dependence as a free parameter when fitting the relation with either calibration sample 
led to similar values of $\beta$$\sim$$-2$ in significant tension with the self-similar value $\beta=-2/3$. We constrained the cosmology with the scaling relations calibrated with a free redshift dependence 
and found essentially identical results with both X-ray calibration samples, with $\sigma_8 = 0.76\pm0.03$, $\Omega_m = 0.34\pm0.03$, and $S_8 \equiv \sigma_8 \sqrt{\Omega_m / 0.3}=0.81\pm0.02$.\\
Compared to results from other probes, they are competitive with analyses of CMB primary anisotropies and cosmic shear in terms of $S_8$ constraining power. Like most analyses of galaxy clusters and other late-time probes, with the exceptions of the XXL survey \citep{garrel_xxl_2022} and the recent eROSITA results presented in \cite{ghirardini_srgerosita_2024}, the cosmological constraints derived in this work show a lower central value of $S_8$ than what is usually found by CMB primary anisotropies analyses. It is important to note that while the central value is lower, none of the four sets of constraints presented here exhibit a strong tension with the latest Planck CMB constraints, with differences ranging from around 2$\sigma$ in the self-similar redshift evolution case and less than 1$\sigma$ when fitting the redshift dependence from the data.\\
This work shows that SZ cluster surveys are a powerful tool to constrain cosmology, and that the mass calibration of the clusters is paramount to the constraining power. In the coming years, the next generation 
of CMB experiments will detect an increasing number of clusters, and a robust mass calibration will be necessary to fully exploit these surveys. Such a mass calibration could be provided by the coming 
weak-lensing and X-ray surveys, that will detect tens of thousands of clusters across a large redshift range.\\

\begin{acknowledgements}
  GA acknowledges financial support from the AMX program. 
  This work was supported by the French Space Agency (CNES).
  This work was supported by the Programme National Cosmology et Galaxies (PNCG) of CNRS/INSU with INP and IN2P3, co-funded by CEA and CNES. 
  This research has made use of the computation facility of the Integrated Data and Operation Center (IDOC, \url{https://idoc.ias.u-psud.fr}) at the Institut d’Astrophysique Spatiale (IAS), 
  as well as the SZ-Cluster Database (\url{https://szdb.osups.universite-paris-saclay.fr}).\\
  WF acknowledges support from the Smithsonian Institution, the \textit{Chandra}
  High-Resolution Camera Project through NASA contract NAS8-0306, NASA
  Grant 80NSSC19K0116 and {\em Chandra} Grant GO1-22132X.\\
  F.A.-S. acknowledges support from {\em Chandra} Grant GO0-21119X.\\
  RJvW acknowledges support from the ERC Starting Grant ClusterWeb 804208.\\
\end{acknowledgements}

\bibliography{bibliotest}
\bibliographystyle{aa}

\begin{appendix}
  \section{Calculation of cluster masses from \textit{Planck} catalogue data}
\begin{figure}[h!]
    \centering
    \includegraphics[width=\columnwidth]{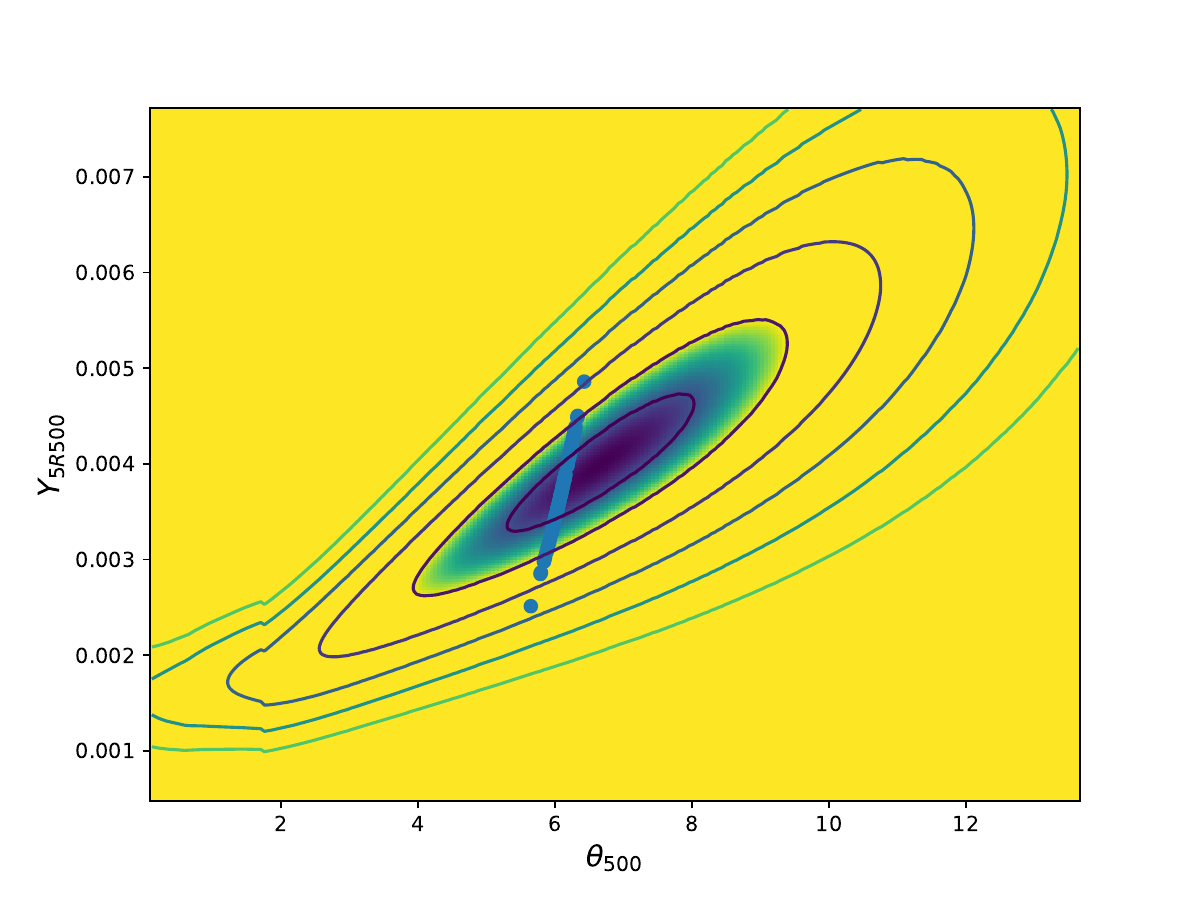}
    \caption{100 samples taken at random from the MCMC mass estimation procedure of the PSZ2 G033.97-76.61 cluster plotted over the 2D probability map from the MMF3 detection found in the PSZ2 catalogue.}
    \label{fig:SZ_mcmc}
\end{figure}

\FloatBarrier
  \label{sz_mcmc}
  The \textit{Planck} MMF3 catalogue provides the 2D posterior probability distribution in $\theta_s$ and $Y_{5R_{500}}$ for each cluster \citep[see][for the full details]{planck_collaboration_planck_2016-1}. 
  Computing masses from these distributions, given a $Y_{\text{SZ}}-M_{500}$ scaling relation, is not straightforward, since the aperture in which the SZ signal is measured depends on $R_{500}$, and thus on the cluster mass.
  To ensure that apertures and masses are coherent, we make use of a second scaling relation between $\theta_{500}$ and $M_{500}$. Unlike the $Y_{\text{SZ}}-M_{500}$ relation, this relation is not calibrated 
  on data but is just a direct consequence of the underlying cosmological model. For the flat $\Lambda$CMD cosmology used in this work, the relation is the following:\\
  \begin{equation}
    \label{eq:theta_500_no_bias}
    \theta_{500}=\theta_{*} \left[\frac{h}{0.7}\right]^{-2/3} \left[\frac{M_{500}}{3.10^{14}M_{\odot}}\right]^{1/3} E^{-2/3}(z) \left[\frac{D_A(z)}{500 \, \text{Mpc}}\right]^{-1},
  \end{equation}
  where $\theta_{*}$=6.997 arcmin.\\
  To obtain masses from the posterior distributions, we use a Monte-Carlo-Markov-Chain (MCMC) approach, using \texttt{emcee} as the sampler. At each step of the chain, we draw values of mass and scaling relation parameters. 
  We then compute the corresponding $\theta_{500}$ and $Y_{\text{SZ}}$, and use them to obtain the likelihood from the observed $\theta_s$-$Y_{5R_{500}}$ probability distribution. We then combine this likelihood 
  with the Gaussian priors on the scaling relation parameters to obtain the final likelihood used to run the chain. The mass is then obtained as the mean of the posterior distribution, with the standard 
  deviation as the uncertainty.
  Fig. \ref{fig:SZ_mcmc} plots 100 samples taken at random from the MCMC mass estimation procedure of the PSZ2 G033.97-76.61 cluster over the 2D probability map found in the PSZ2 catalogue.
\FloatBarrier
  \section{Importance of proper non-detection treatment in bias estimations}
  \label{selection_effects}
\begin{figure}[h!]
    \centering
    \includegraphics[width=\columnwidth]{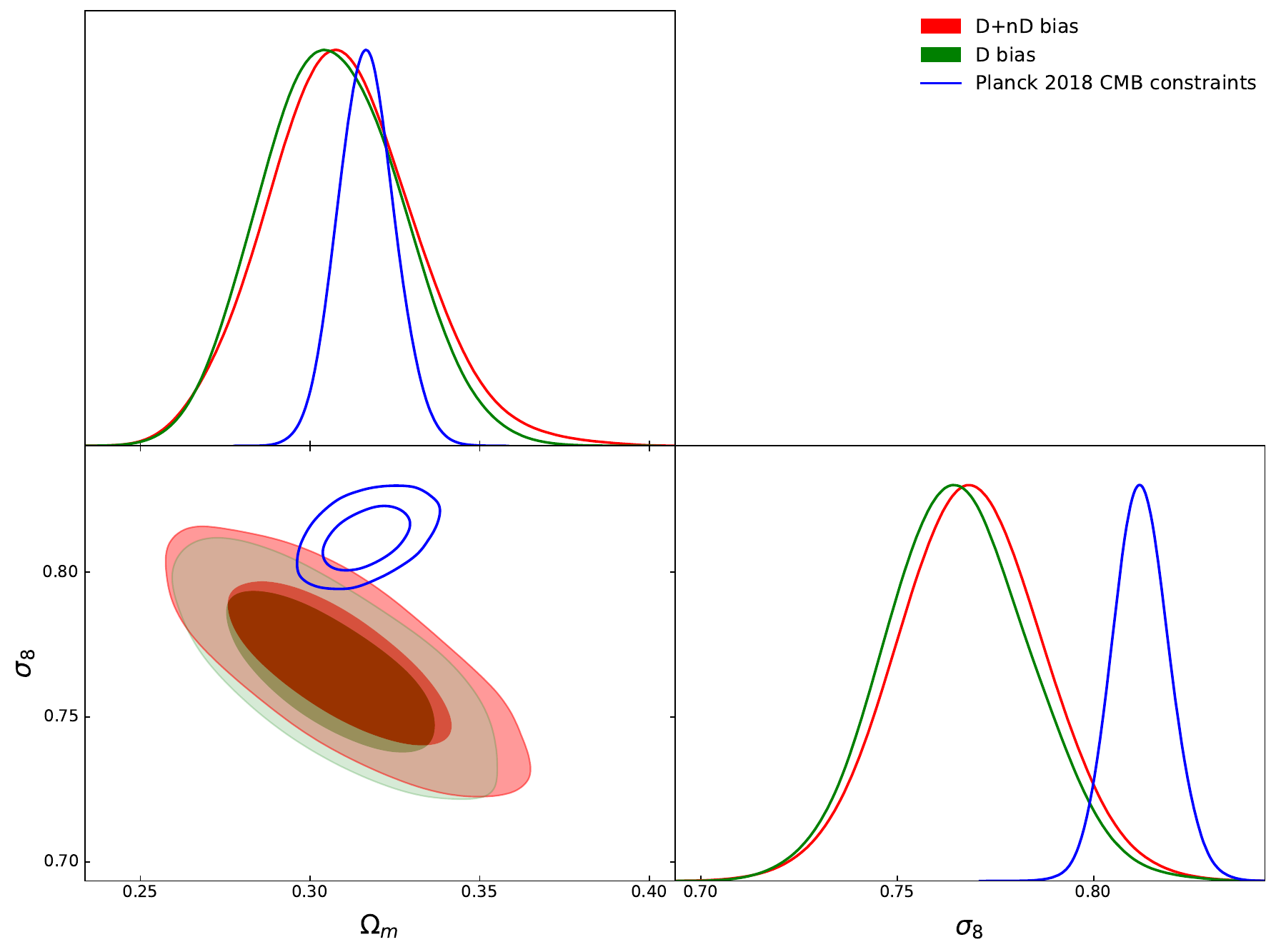}
    \caption{Final cosmological constraints obtained with the bias calculated not including the non-detections (in green) or taking them into account (in red).}
    \label{fig:constraints_bias}
\end{figure}

  Fig. \ref{fig:constraints_bias} shows the difference in the final cosmological constraints with the same scaling relation, but using either the bias calibrated using only the cluster detected 
  by \textit{Planck} (D bias) or also accounting for non-detections (D+nD bias). The shift in the constraints is small, with a preference for slightly higher values of $\sigma_8$ and marginally better constraining power 
  when accounting for non-detections: $\sigma_8^{\text{D+nD}}=0.765\pm0.019, \, \sigma_8^{\text{D}}=0.760\pm0.020$. The value and uncertainty of $\Omega_{\text{m}}$ are not affected.
\FloatBarrier
  \section{Synthetic sample generation}
  \label{toy_model}
\begin{figure}[h!]
    \centering
    \includegraphics[width=\columnwidth]{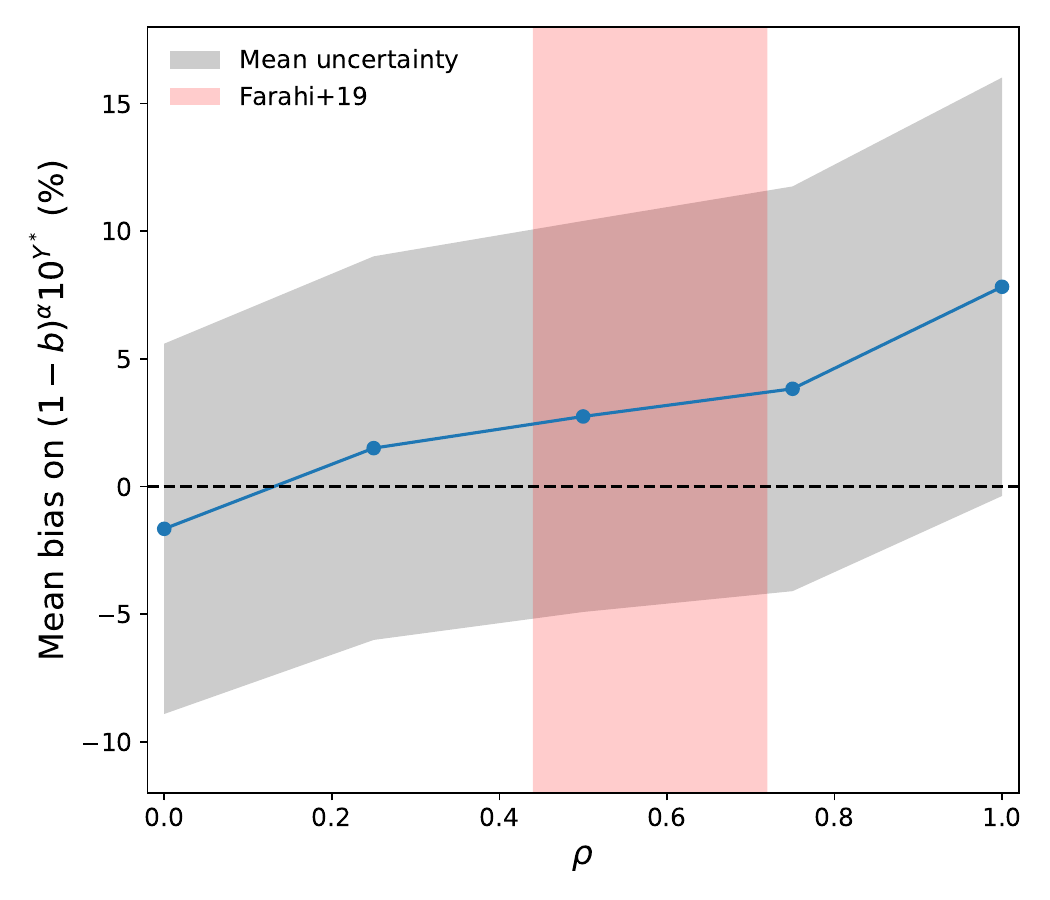}
    \caption{Evolution of the mean bias on the parameter combination $(1-b)^{\alpha}10^{Y^*}$. The grey contours represent the mean uncertainty on the parameter combination returned by our scaling relation calibration pipeline, and the red contours show the $\rho$ estimate obtained by \cite{farahi_detection_2019}.}
    \label{fig:toy_model_correlated}
\end{figure}

  In order to verify that our mass calibration pipeline retrieves the correct underlying scaling relation, we generate realistic mock calibration samples with a known underlying scaling relation and run them through the scaling relation calibration pipeline. We first create a 2D number counts array by Poisson sampling the \cite{tinker_toward_2008} halo mass function, calculated using the \texttt{hmf} Python package \citep{murray_hmfcalc_2013}, in 350 redshift bins between $z=0$ and $z=0.5$ and 500 $M_{500}$ bins between $10^{14}$ and $10^{16} M_\odot$. We attribute to each cluster a true X-ray mass,
  \begin{equation}
    \label{eq:m_x_toymodel}
    M_{500}^{\text{X, true}}=(1-b)M_{500}e^{x_1},
  \end{equation}
  a true weak-lensing mass, and
  \begin{equation}
    \label{eq:m_wl_toymodel}
    M_{500}^{\text{WL, true}}=M_{500}e^{x_2},
  \end{equation}
  a true SZ-signal,
  \begin{equation}
    \label{eq:ysz_toymodel}
    Y_{\text{SZ, true}}=Y_{\text{piv}}\frac{E^{2/3}(z) 10^{Y^*}}{D^2_A}\left(\frac{M_{500}^{Y_{\text{X}}}}{M_{\text{piv}}}\right)^{\alpha}e^{x_3},
  \end{equation} 
  where the $x_i$ variables are used to generate random log-normal intrinsic scatter, with the following:
  \begin{equation}
    \label{eq:int_scatter_toymodel}
     P(x_i) = \mathcal{N}\left(\begin{bmatrix} 0 \\ 0 \\ 0 \end{bmatrix} \Sigma \right) \text{, where } \Sigma = \begin{bmatrix} \sigma^2_\text{X} & 0 & \rho \sigma_\text{X} \sigma_\text{SZ} \\ 0 & \sigma_\text{WL}^2 & 0 \\ \rho \sigma_\text{X} \sigma_\text{SZ} & 0 & \sigma^2_\text{SZ}\end{bmatrix}.
  \end{equation}
  We note that we allow for correlated intrinsic scatter between X-ray mass and SZ-signal, since the X-ray masses in the real data are derived from $Y_\text{X}$ which might show correlation with $Y_\text{SZ}$. In the main body, we presented results for no correlation, but we conducted the study for $\rho=[0, 0.25, 0.5, 0.75, 1]$ to quantify the effect of this potential correlation.\\
  We use a simple neural network (two-layer perceptron with 1000 units each) trained on the actual data to generate realistic uncertainties on the X-ray masses given the X-ray mass and redshift. We multiply the output of the network by a random log-normal component to add randomness to the uncertainties to best match the actual data. We repeat the same procedure to generate uncertainties on the WL masses. We then obtain measured X-ray and WL masses $M_{500}^{\text{X, mes}}$ and $M_{500}^{\text{WL, mes}}$ by adding a random Gaussian component, with mean 0 and standard deviation the uncertainty previously generated, to $M_{500}^{\text{X, true}}$ and $M_{500}^{\text{WL, true}}$.\\
  We attribute a random position on the sky to each cluster, from which we determine the true S/N of the cluster based on the Planck SZ noise maps. The measured S/N is obtained by adding a random Gaussian component of mean 0 and variance 1, and the measured SZ signal and uncertainty are directly derived from the measured S/N and noise.\\
  We create the mock X-ray calibration sample by compiling all the clusters with $z<0.35$, a position in the unmasked regions of the Planck sky and a S/N above 7.5. This slightly higher S/N cut compared to the ESZ sample is due to the fact that the noise maps used are DR2 maps, which are less noisy than the early maps used to derive the ESZ sample. Since we cannot accurately model the weak-lensing selection function, we create the mock weak-lensing sample by randomly selecting 56 clusters outside of the masked regions with and a SZ S/N>6 to mimick the 56 PSZ2 clusters with WL data and 41 clusters outside of the masked regions with a S/N below 6 to mimick the clusters with WL masses that are not detected by Planck.\\
  We repeat the sample generation procedure 100 times for each value of $\rho$, running the scaling relation fitting pipeline on the synthetic samples every time to ensure that the underlying scaling relation is properly recovered. As shown in Fig. \ref{fig:toy_model_res}, the scaling relation is retrieved without any significant bias when $\rho=0$. When correlation is introduced, some bias starts to appear. In the case of the slope $\alpha$ and intrinsic scatter $\sigma_\text{int}$, the bias remains small enough that the effect is completely negligible given the low sensitivity of the final constraints on these particular parameters. Fig. \ref{fig:toy_model_correlated} presents the evolution of the bias on the parameter combination $(1-b)^{\alpha}10^{Y^*}$, which sets the mass scale and, as such, has an important effect on the final constraints.
  While the bias grows larger as the correlation increases, it remains below 1$\sigma$ even in the $\rho=1$ case. \cite{farahi_detection_2019} studied this correlation using the Local Cluster Substructure Survey data, and found a correlation coefficient $\rho=0.60^{+0.12}_{-0.16}$. If we combine this correlation estimate with our study of synthetic cluster samples, we find that the value of $(1-b)^{\alpha}10^{Y^*}$ retrieved by our method is not significantly biased, with a mean bias of $\sim$3\%, or about 0.4$\sigma$. This $(1-b)^{\alpha}10^{Y^*}$ shift of 3\% is expected to lead to a final shift of $S_8$ of less than 0.5\%, thus being negligible given the constraining power of our results.
\FloatBarrier
  \section{Representativity of the calibration samples}
  \label{representativity}
\begin{figure}[h!]
    \centering
    \includegraphics[width=\columnwidth]{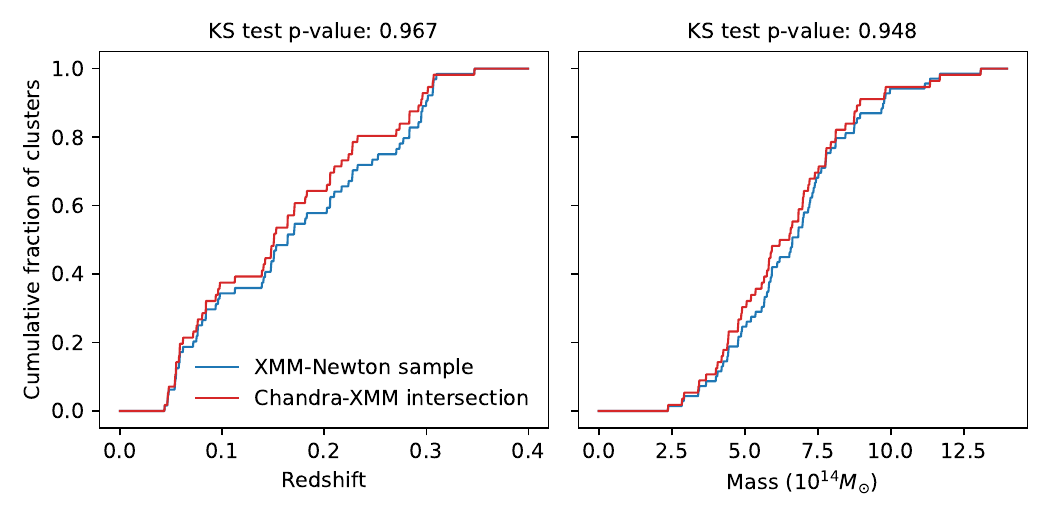}
    \caption{Comparison of the redshift and mass distribution of the \textit{XMM-Newton} sample (in blue) with the \textit{Chandra}-\textit{XMM-Newton} intersection sample (in red). The p-value obtained when performing a KS-test comparing the two distributions is reported above each plot.}
      \label{fig:Mz_dist_XMM_Chandra}
    \end{figure}

\begin{figure}[h!]
    \centering
    \includegraphics[width=\columnwidth]{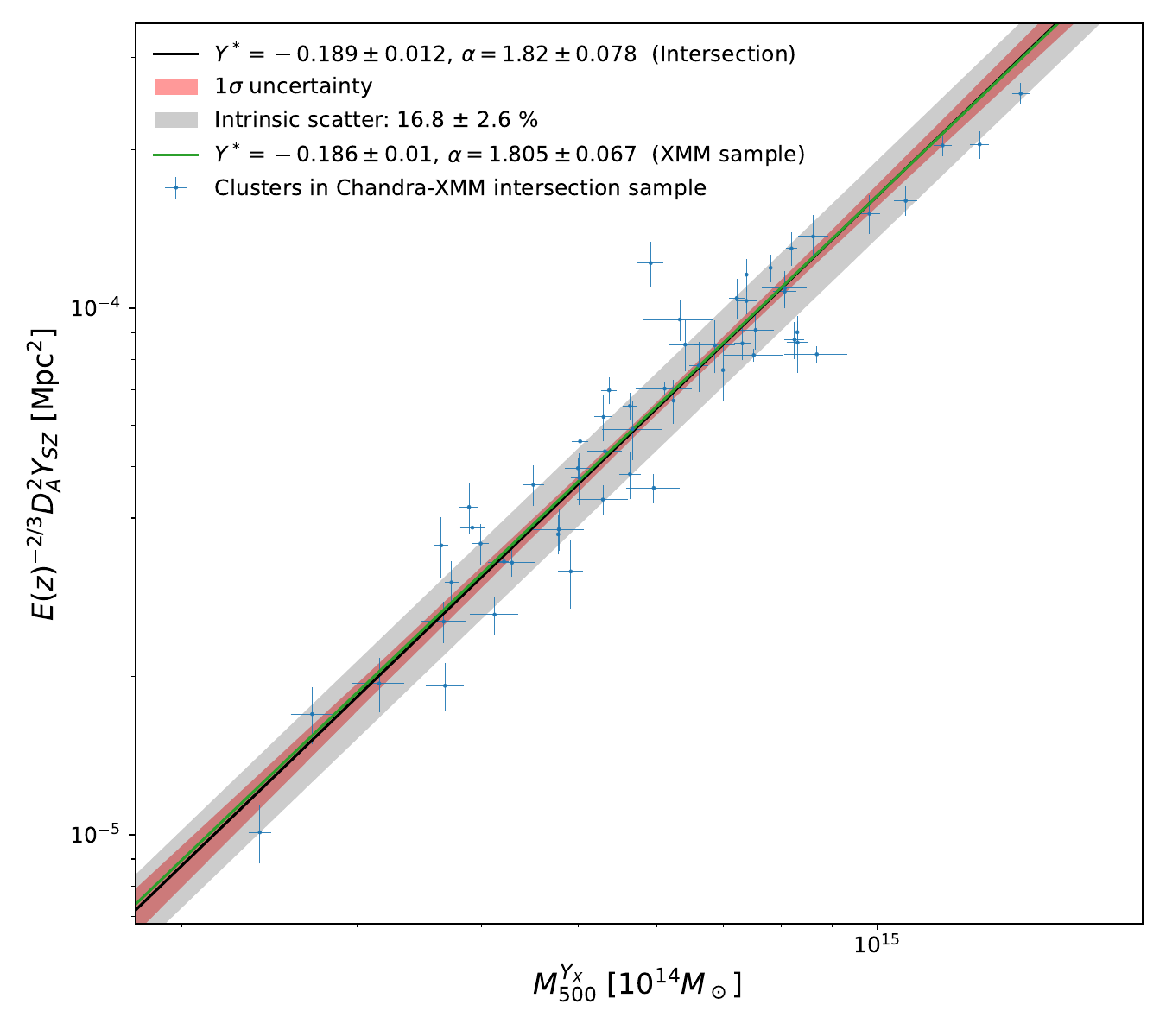}
    \caption{Comparison of the scaling relation obtained with the \textit{XMM-Newton} sample restricted to the intersection (black line with uncertainty and intrinsic scatter) and the relation obtained with the full \textit{XMM-Newton} sample (green line).}
      \label{fig:Scaling_intersection}
    \end{figure}

  The \textit{Chandra-Planck} sample used in this work contains 56 of the 71 clusters of the \textit{XMM-Newton}-sample used by the \textit{Planck} collaboration. Since six of the missing clusters are above the redshift limit of the \textit{Chandra} sample, 
  56 of the 65 clusters of the \textit{XMM-Newton}-sample that are below the redshift limit of the \textit{Chandra} sample are included in the \textit{Chandra-Planck} sample. To verify that the subset of the \textit{XMM-Newton}-sample 
  contained in the \textit{Chandra} sample is representative of the full \textit{XMM-Newton} sample, we compare mass and redshift distributions via Kolmogorov–Smirnov (KS) tests. We use \textit{XMM-Newton} masses for all clusters and restrict the 
  \textit{XMM-Newton} sample to z<0.35 when comparing redshift distributions to not introduce a bias due to different ranges in the KS-test. Fig. \ref{fig:Mz_dist_XMM_Chandra} presents the results of the KS tests: no indication of a different mass or redshift distribution is found. A final check 
  is to calibrate the $Y_{\text{SZ}}-M_{500}^{Y_{\text{X}}}$ scaling relation with either the full \textit{XMM-Newton} sample \citep[expecting the same result as in][]{planck_collaboration_planck_2014} or with the \textit{XMM-Newton} sample restricted 
  to clusters present in the \textit{Chandra-Planck} sample. Fig. \ref{fig:Scaling_intersection} shows the scaling relations obtained for both samples. The difference in both intercept and slope is not statistically significant at around 0.3$\sigma$, 
  meaning that the intersection sub-sample is not biased with respect to the full \textit{XMM-Newton} sample in the $Y_{\text{SZ}}-M_{500}^{Y_{\text{X}}}$ plane. \textit{XMM-Newton} derived masses were used in both cases and no redshift cut 
  was applied when performing this comparison.\\
  \\
  Given the above results of the comparison between the intersection of the \textit{Chandra-Planck} and \textit{XMM-Newton} samples and the \textit{XMM-Newton} sample, we conclude that the intersection is representative of the full \textit{XMM-Newton} sample, 
  and that we can thus compare the intersection with the \textit{Chandra} sample and extrapolate the results to the full \textit{XMM-Newton} sample. This extrapolation is required as we need to have X-ray data from the same instrument, 
  in order to be able to draw conclusions from sample comparisons.\\
  To assess whether the \textit{XMM-Newton} calibration sample used in \cite{planck_collaboration_planck_2014} and \cite{planck_collaboration_planck_2016} is representative of the cosmological sample on which the scaling relation 
  is applied, we first compare the mass and redshift distributions of the \textit{XMM-Newton} calibration sample, the \textit{Chandra} calibration sample, and the full \textit{Planck} cosmological sample. Given the limitations of the KS-test, 
  we apply a z<0.35 cut when comparing redshift distributions. Since we want to compare how representative of the full cosmology sample the calibration samples are, no redshift cut is applied when studying mass distributions.
  \begin{figure*}
    \centering
    \resizebox{0.8\hsize}{!}
          {\includegraphics{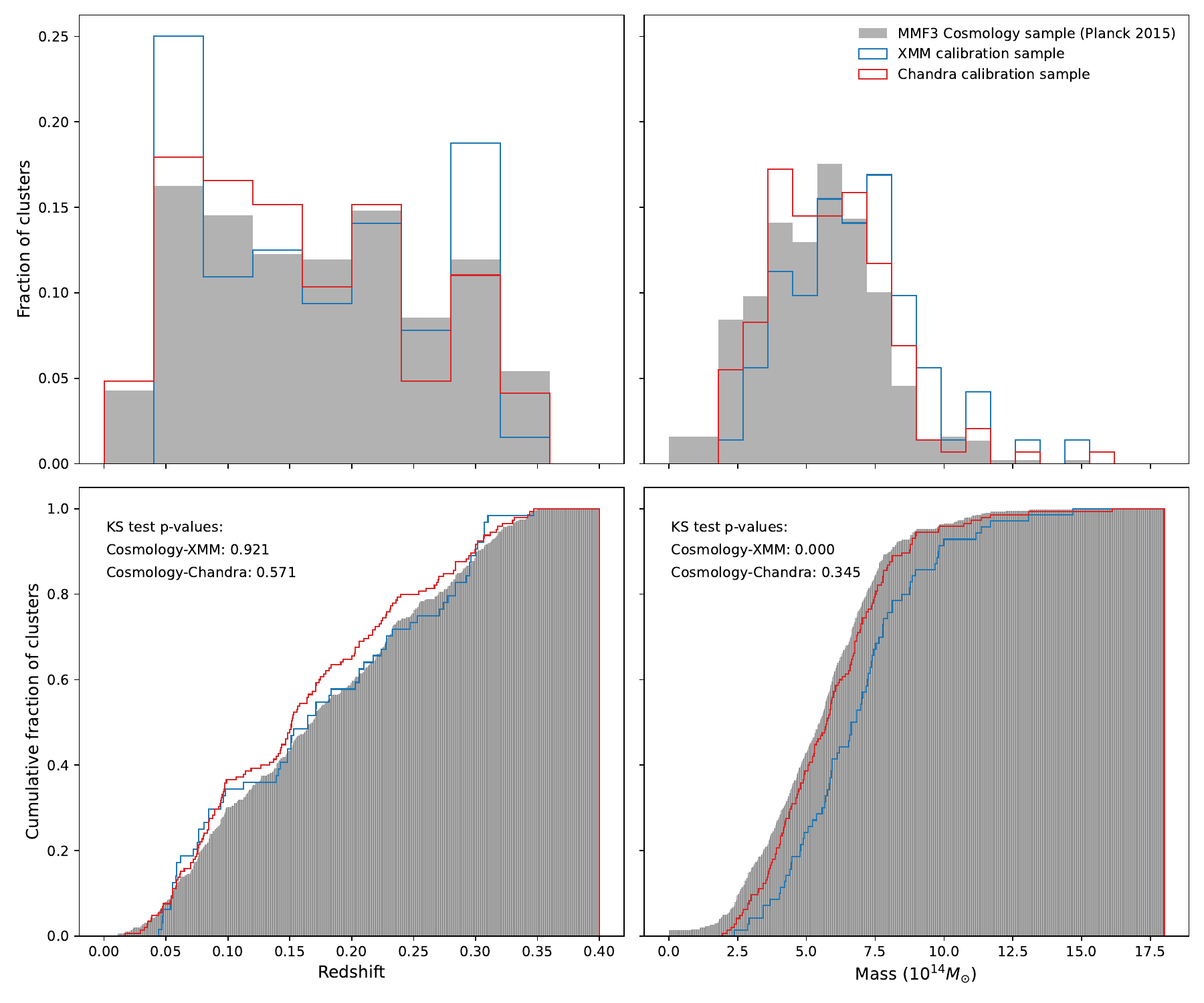}}
    \caption{Comparison of the redshift (left side) and mass (right side) distributions of the \textit{XMM-Newton} sample (in blue) and the \textit{Chandra} sample (in red) with the \textit{Planck} cosmology sample. 
              The p-values obtained when performing a KS-test comparing each calibration sample with the cosmology sample is reported in the bottom panels.}
        \label{fig:cosmo_calibration}
  \end{figure*}
  \begin{figure*}
    \centering
    \resizebox{0.99\hsize}{!}
          {\includegraphics{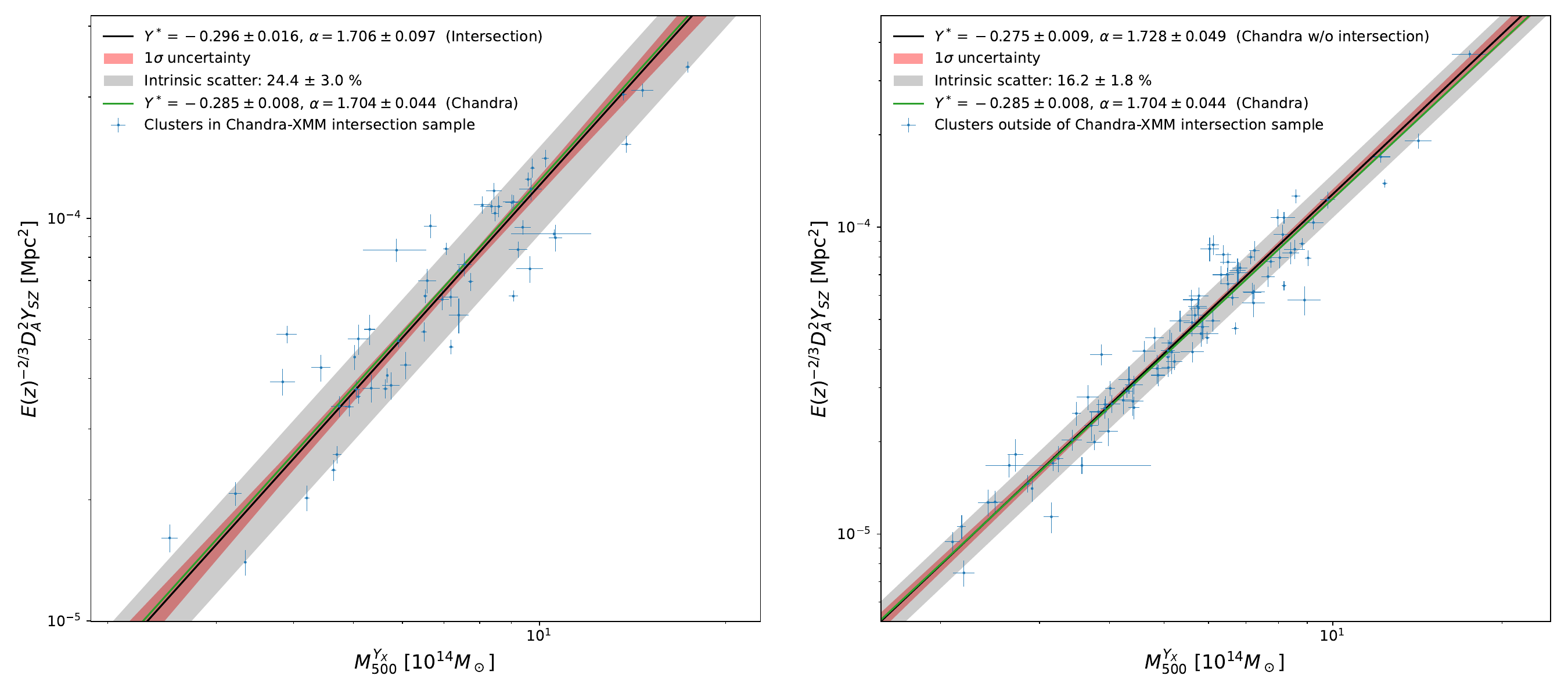}}
    \caption{Scaling relations obtained when including or excluding clusters of the \textit{XMM-Newton} sample.
      \textbf{Left:} Comparison of the scaling relation obtained with the \textit{Chandra} sample restricted to the 56 clusters present in both \textit{Chandra} and \textit{XMM-Newton} samples 
    (black line with uncertainty and intrinsic scatter) and the relation obtained with the full \textit{Chandra} sample (green line).\\
    \textbf{Right:} Comparison of the scaling relation obtained with the \textit{Chandra} sample restricted to the 110 clusters present in the \textit{Chandra} sample and absent from the \textit{XMM-Newton} sample 
    (black line with uncertainty and intrinsic scatter) and the relation obtained with the full \textit{Chandra} sample (green line).}
        \label{fig:representativity}
  \end{figure*}
  
  Fig. \ref{fig:cosmo_calibration} shows the results of the comparisons. The KS-tests do not point to significant differences between the redshift distributions of the calibration samples and the cosmology sample, 
  although the \textit{Chandra} calibration sample contains more low-redshift clusters even after restricting all samples to z<0.35. In terms of mass distribution, the comparisons show 
  that both calibration samples are biased towards higher masses than the cosmology sample, with the bias being much stronger in the case of the \textit{XMM-Newton} sample (p-value<0.0005 vs p-value=0.345). 
  In the case of the \textit{Chandra-Planck} sample, both of those differences are expected, since the ESZ and Cosmology samples are signal-to-noise limited with the same cut, but the ESZ sample was obtained on a noisier map, 
  therefore requiring higher signal and ultimately more massive/less distant clusters. The bias of the \textit{XMM-Newton} sample towards higher masses is also expected, since it is a compilation of pre-existing \textit{XMM-Newton} observations, 
  which are more likely to be targeted towards bright and massive clusters, as well as having a higher SZ signal-to-noise cut (7 as opposed to 6 for the cosmology sample).\\
  The next, and arguably more important test, is to see if the \textit{XMM-Newton} calibration sample is biased in the $Y_{\text{SZ}}-M_{500}^{Y_{\text{X}}}$ plane compared to an SZ-selected sample. Fig. \ref{fig:representativity} shows 
  the scaling relation obtained with the \textit{Chandra} calibration sample restricted to the 56 clusters present in both \textit{Chandra} and \textit{XMM-Newton} samples (left panel), and the relation obtained with the \textit{Chandra} calibration sample 
  restricted to the 110 clusters present in the \textit{Chandra} sample and absent from the \textit{XMM-Newton} sample (right panel), and compares them to the scaling relation calibrated in this work on the full \textit{Chandra} sample.
  As previously discussed, we need to use the intersection of the \textit{Chandra} and \textit{XMM-Newton} samples instead of the full \textit{XMM-Newton} sample, since we need to have X-ray data from the same instrument in order to be able to draw conclusions from sample comparisons.
  We showed in the previous paragraph that the intersection of the \textit{Chandra} and \textit{XMM-Newton} samples is representative of the full \textit{XMM-Newton} sample, and we can thus extrapolate the results obtained with the intersection to the full \textit{XMM-Newton} sample.
  Although not very significant (respectively 0.7$\sigma$ and 1$\sigma$), the scaling relation obtained with the intersection of the \textit{Chandra} and \textit{XMM-Newton} samples (respectively the exclusion of the \textit{XMM-Newton} sample) has a lower 
  (respectively higher) normalisation than the one obtained with the full \textit{Chandra} sample. This means that, as could be expected, the \textit{XMM-Newton} sample is slightly biased towards X-ray bright clusters, when compared to the SZ-selected \textit{Chandra} sample.
  \FloatBarrier
  \section{Comparison of MMF algorithms}
  \label{MMF_comparison}
\begin{figure}[h!]
    \centering
    \includegraphics[width=\columnwidth]{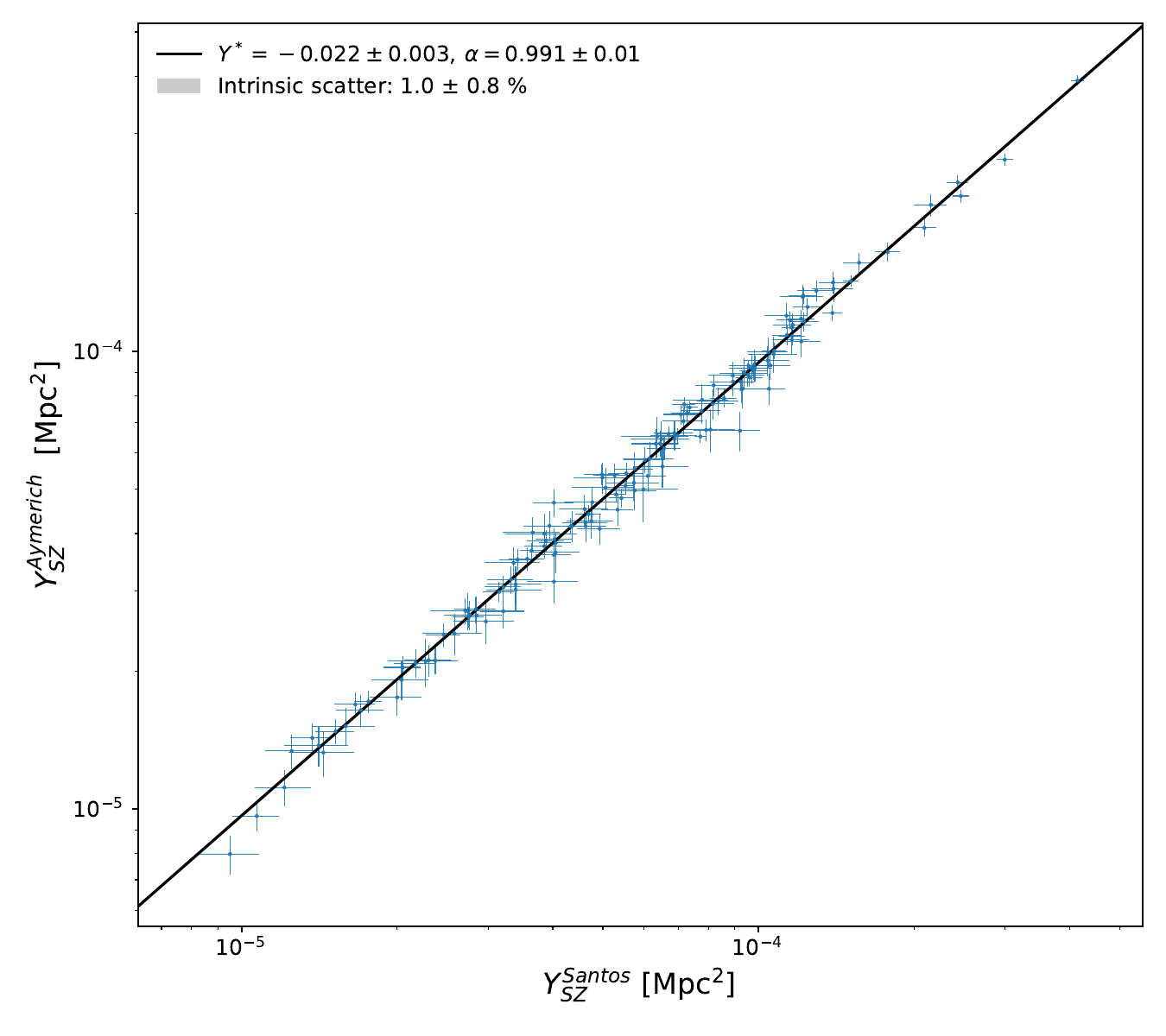}
    \caption{Comparison of $Y_{\text{SZ}}$ obtained in this work and in Andrade-Santos et al. 2021, with the same X-ray derived positions and aperture as input, but independently developed MMF methods.}
    \label{fig:mmfs}
  \end{figure}

  As mentioned in Sect. \ref{SZ_data}, the extraction of the SZ signal from the \textit{Planck} maps using fixed positions and apertures from the X-ray data was done in \cite{andrade-santos_chandra_2021} and in this work 
  using fully independently developed MMF algorithms. Fig. \ref{fig:mmfs} shows the comparison of the $Y_{\text{SZ}}$ measurements obtained in both works. The agreement is very good, with no departure from unity in 
  both the slope and intercept of the best-fit relation. The intrinsic scatter around the best fit is also very low at 1\%. For reference, this intrinsic scatter is better than the one obtained when comparing the 
  results of MMF1 and MMF3 algorithms in \cite{planck_collaboration_planck_2016-1}, which is around 3\%. We are therefore confident that the $Y_{\text{SZ}}$ measurements obtained in this work are robust.
\FloatBarrier
\newpage
  \section{Full triangle plot of MCMC fitting}
\begin{figure*}
    \centering
    \resizebox{0.99\hsize}{!}
          {\includegraphics{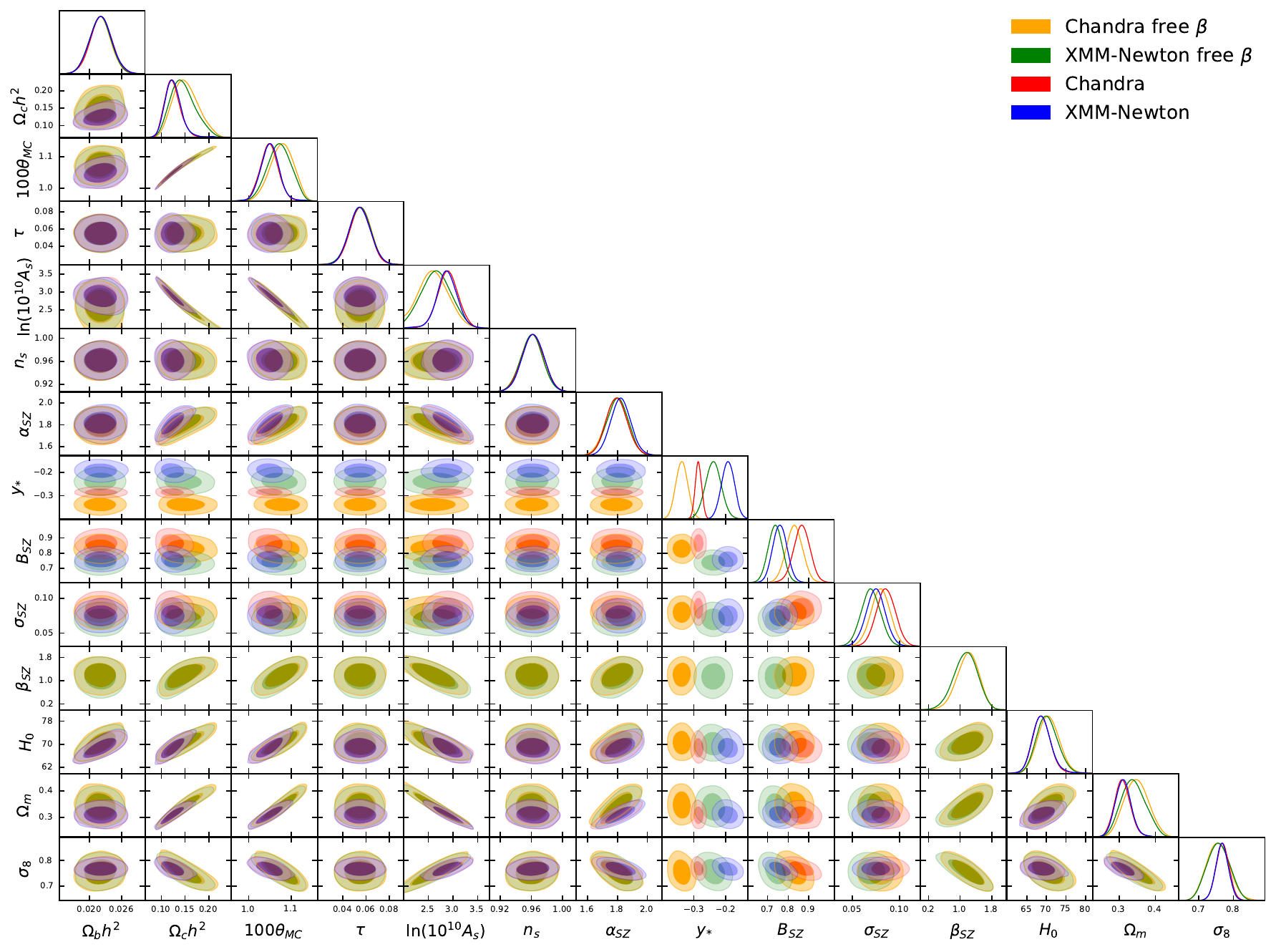}}
    \caption{Parameter constraints obtained from the MCMC fitting of the number counts, for all cases considered in this work.}
        \label{fig:full_triangle}
  \end{figure*}

  Fig. \ref{fig:full_triangle} shows the full parameter constraints of the MCMC fitting of the number counts.
  \FloatBarrier
  \end{appendix}

\end{document}